\documentclass[11pt,a4paper]{article}
\usepackage[left=20mm,top=20mm,right=20mm,bottom=18mm]{geometry}
\usepackage{longtable}
\usepackage{amsmath, amsthm, amsfonts}
\usepackage{natbib}
\usepackage[colorlinks=true,urlcolor=blue,citecolor=blue]{hyperref}
\usepackage{url}
\newcounter{bibcount}
\makeatletter
\patchcmd{\@lbibitem}{\item[}{\item[\hfil\stepcounter{bibcount}{\thebibcount.}}{}{}
\renewcommand\NAT@bibsetup%
   [1]{\setlength{\leftmargin}{\bibhang}\setlength{\itemindent}{-\parindent}%
       \setlength{\itemsep}{\bibsep}\setlength{\parsep}{\z@}}
\makeatother
\setcitestyle{authoryear,open={(},close={)}}
\newcommand{\textcite}[1]{\cite{#1}}

\usepackage[colorlinks=true,urlcolor=blue,citecolor=blue]{hyperref}
\usepackage{overpic}
\usepackage{relsize}
\usepackage{booktabs}
\usepackage{pdflscape}
\usepackage{multirow}
\usepackage{caption}
\allowdisplaybreaks
\usepackage{xcolor}

\newcommand{\black}[1]{{\color{black} #1}}
\newcommand{\blt}{\noindent $\bullet$}

\usepackage[scaled]{helvet}
\makeatletter
\renewcommand{\paragraph}{%
  \@startsection{paragraph}%
  {4}%    % level
  {\z@}%  % indentation
  {0.5ex \@plus.2ex \@minus.2ex}%  % before skip (adjustable)
  {-1em}% % after skip
  {\normalfont\normalsize\itshape}%  % style
}
\makeatother

\usepackage{comment}
\excludecomment{journal}
\includecomment{plain}

\newcommand{\rev}[1]{{#1}}
\begin{document}

% title
\title{Dominant ionic currents in rabbit ventricular action potential dynamics}
\author{Zhechao Yang $^{1}$, Hao Gao $^{1}$,  Godfrey L Smith $^{2}$ and Radostin D Simitev\,$^{1,\ast}$ \\[5mm]
  {\small
    $^{1}$ School of Mathematics \& Statistics, University of
    Glasgow, Glasgow, UK} \\
  {\small
    $^{2}$  School of Cardiovascular \& Metabolic Health University of Glasgow, Glasgow, UK} \\  
  {\small  $^\ast$ Corresponding author:
    \href{mailto:Radostin.Simitev@glasgow.ac.uk}{Radostin.Simitev@glasgow.ac.uk},
    \href{https://orcid.org/0000-0002-2207-5789}{orcid.org/0000-0002-2207-5789}}}
\date{\today}
\maketitle

%%%% Abstract text to be placed here %%%%%%%%%%%%
\begin{abstract}
  Mathematical models of cardiac cell electrical activity include 
  numerous parameters, making calibration to experimental data and
  individual-specific modeling challenging. This study applies Sobol
  sensitivity analysis, a global variance-decomposition method, to identify
  the most influential parameters in the Shannon model of rabbit
  ventricular myocyte action potential (AP). The analysis highlights
  the background chloride current ($I_\text{Clb}$) as the dominant determinant
  of AP variability. Additionally, the inward rectifier potassium
  current ($I_\text{K1}$), fast/slow delayed rectifier potassium
  currents ($I_\text{Kr}$, $I_\text{Ks}$), sodium-calcium exchanger
  current ($I_\text{NaCa}$), the slow component of the transient
  outward potassium current ($I_\text{tos}$), and L-type calcium  
  current ($I_\text{CaL}$) significantly affect AP biomarkers, including
  duration, plateau potential, and resting potential. Exploiting these
  results, a   hierarchical reduction of the model is performed and
  demonstrates that retaining only six key   parameters can
  sufficiently   capture individual biomarkers, with a   coefficient
  of determination   exceeding 0.9 for selected   cases. These
  findings improve the utility of the   Shannon model   for
  personalized simulations, aiding   applications like   digital
  twins and drug response predictions in   biomedical research. 

\end{abstract}
%%%%%%%%%%%%%%%%%%%%%%%%%%%                        

\subsection*{Author Summary}
\begin{quote}
In our study, we explored the complexity of a mathematical model used
to understand the electrical activity of rabbit ventricular
cells. Such models involve many parameters, which makes it difficult
to match them with experimental data or to personalize them for
individual cells. To address this, we used a method for global
sensitivity analysis to identify which ion current parameters have the
most impact on the action potential output of the model.
Our analysis shows that the background chloride current is the main
factor influencing the variability of the rabbit ventricular action
potential. Other important currents include the inward rectifier
potassium current, various potassium currents, sodium-calcium
exchanger current, the slow component of the transient outward
potassium current, and the L-type calcium current. These factors play
a significant role in determining key features of the action
potential, such as its duration, plateau potential, and resting potential.
Based on these results, we proposed a sequence simplified models
obtained by retaining variability in a few most influential parameters
while fixing the rest to appropriate constant values. This approach
is also applicable to other mathematical models of cardiac action
potentials and can make them more useful for personalized simulations,
and in areas like digital twins and predicting drug responses in
biomedical research. 
\end{quote}

\section{Introduction}

% State what this paper is about
Mathematical models of cardiac action potentials (APs) are essential
tools for studying the electrical activity of cardiac cells and
understanding their physiological and pathological behaviors. 
In this work, we conduct a global sensitivity analysis of a
mathematical model of rabbit ventricular myocyte APs, and
rank key parameters based on their impact on aspects of the
model's output.  

% Clarify various terms and concepts:
%  - APs     
Ventricular APs are transient deviations from the
resting electrical potential across the sarcolemma, and play an
essential role in cardiac function. They are triggered by electrical
stimuli and dynamically sustained by ionic conductance changes
regulated by voltage and ion concentrations.   
%  - Mathematical models of the AP
The complexity of the underlying cellular processes and
structures necessitates nonlinear, stiff, and high dimensional
mathematical models of the AP \citep{Amuzescu2021}. Consequently, their analysis,
numerical simulation, and comparison with experimental data is 
challenging. Often a model reduction is needed to a simplified system
while retaining the predictive power of the original model.  

% - sensitiivity analysis
A widely used approach for model reduction and simplification is
sensitivity analysis. This method can be viewed as an optimization
problem, where the objective is to minimize the model's dimensionality
while ensuring that the discrepancy between the outputs of the
original and reduced models remains sufficiently small. In practice,
this involves identifying model quantities whose variations have
negligible impact on the output and replacing them with appropriate
constants. 
% advantages of sensitivity analysis to other methods
Compared to other model reduction techniques such as variable lumping,
coordinate transformations, manifold reduction, truncation via
singular value decomposition, or homogenization, sensitivity analysis
offers a key advantage: it retains the physiological interpretability
of model variables, parameters, and components
\citep{Snowden2017}.
% This makes it particularly valuable for calibrating detailed
% mechanistic models to experimental data, as required in applications
% like digital twinning and patient-specific predictions. 

% Specific motivation
This study is motivated by the need to model recent experimental
measurements reported by \cite{Lachaud2022}. In that
work, AP waveforms were measured in hundreds of
isolated rabbit ventricular myocytes, both before and after drug
administration. Cells were sourced from various regions of the
left ventricle across multiple animals. The most surprising finding
was that AP waveforms varied more significantly between individual
cells than between regions or even between different animals. 
Cell-specific mathematical models are required to capture this
pronounced cell-to-cell variability. 
Such models can be developed by adjusting a subset of parameters in a
generic baseline model, such as \citep{shannon2004mathematical}.
To achieve this, a critical step is understanding how variations in
parameter values influence the AP waveform. The overall goal of this
study is to perform this sensitivity analysis, with specific
objectives outlined further below.

\newcommand{\customsize}{\fontsize{8.1pt}{9.9pt}\selectfont}

\renewcommand{\arraystretch}{1.2}
\begin{longtable}[t]{p{0.3\textwidth}p{0.3\textwidth}p{0.3\textwidth}}
\label{tab:literature} \\
\toprule
  Goals & Methods \& Models & Results \\
\midrule
\endfirsthead

\multicolumn{3}{l}{{\customsize \textit{Continued from previous page}}} \\
\toprule
  Goals & Methods \& Models & Results \\
\midrule
\endhead

\midrule
\multicolumn{3}{r}{{\customsize \textit{Continued on next page}}} \\
\endfoot

\bottomrule \\
\caption{Selected works on sensitivity analysis of cardiac cellular electrophysiology models}
\endlastfoot

\multicolumn{3}{l}{\citet{gemmell2014population}} \\
\blt To perform a systematic exploration of the effects
of simultaneously varying the magnitude of six transmembrane current conductances in two
rabbit-specific ventricular AP models.
& \blt Clutter-based dimension reordering. \blt Rabbit-specific ventricular AP
model: \citet{shannon2004mathematical}. 
& \blt $g_\text{Ca,L}$ had the
greatest influence on AP duration variability at both 400 and 1000 ms, along with $g_\text{K1}$ and $g_\text{to}$ at 400 and 1000 ms, respectively.\\
\midrule

\multicolumn{3}{l}{\citet{sobie2009parameter} } \\
\blt To investigate how maximum ion channel conductances affect the outputs of a simplified cardiac cell model, specifically focusing on AP shape and restitution.
& \blt Multivariable Regression. \blt Cardiac cell models: \citet{luo1991model}, \citet{fox2002ionic}, \citet{kurata2005dynamical}. 
& \blt Blocking rapid delayed rectifier current ($I_\text{Kr}$) causes a greater prolongation of AP duration than blocking slow delayed rectifier current ($I_\text{Ks}$), which has little effect. Changes in inward rectifier current ($I_\text{K1}$) have a much greater effect on AP duration than changes in $I_\text{Kr}$ in the model of \citet{fox2002ionic}.\\
\midrule

\multicolumn{3}{l}{\citet{coveney2020sensitivity} } \\
\blt To study how different ionic conductances in two cardiac cell models affect the AP duration and other electrophysiological properties.
& \blt Variance based sensitivity analysis. \blt Cardiac cell models: \citet{courtemanche1998ionic}, \citet{maleckar2009k+}. 
& \blt Changes in maximum conductance of the ultra-rapid K + channel ($I_\text{Kur}$) would have opposite effects on AP duration.\\
\midrule

\multicolumn{3}{l}{\citet{romero2009impact} } \\
\blt To investigate the impact of variability in ionic currents on the electrophysiological properties of human ventricular cells.
& \blt Local sensitivity analysis. \blt Human ventricular model: \citet{ten2006alternans}. 
& \blt AP duration is moderately sensitive to changes in maximal conductances of all currents involved in repolarization and also $I_\text{Ks}$ and $I_\text{CaL}$ kinetics.\\
\midrule

\multicolumn{3}{l}{\citet{chang2015bayesian}} \\
\blt To investigate how uncertainty in input parameters influences electrophysiological features and mechanical variables.
& \blt Variance based sensitivity analysis. \blt Cardiac cell model: \citet{luo1991model}. 
& \blt The analysis showed that parameters like $G_{\text{si}}$, $G_{\text{K}}$, and $G_{\text{b}}$ were crucial in determining the variability in AP duration, a key electrophysiological measure.\\
\midrule

\multicolumn{3}{l}{\citet{romero2011systematic}  } \\
 \blt To characterise the sensitivity of selected cellular
         biomarkers of arrhythmic risk to ionic current properties in rabbit
         ventricular models. \blt To compare with experimental data.&\blt Local sensitivity analysis.\blt Rabbit cell models: \citet{shannon2004mathematical}, \citet{Mahajan2008}.&\blt AP duration is significantly influenced by most repolarization currents. \blt APtriangulation is mainly regulated by $I_\text{K1}$.\blt AP duration restitution properties, and Ca$^{2+}_i$ rate dependence arestrongly affected by $I_\text{NaK}$, $I_\text{CaL}$ and $I_\text{NaCa}$.\\     
\midrule

\multicolumn{3}{l}{\citet{johnstone2016uncertainty}  } \\
\blt To investigate the impact of variability in ionic currents on the electrophysiological properties of human ventricular cells.
& \blt Variance based sensitivity analysis. \blt Human ventricular model: \citet{ten2006alternans}. 
& \blt $V_\text{max}$ is sensitive to $G_\text{Na}$, $G_\text{CaL}$. $V_\text{m}$, $A_{90}$ and $A_{50}$ are both influenced by a similar group of inputs, and resting voltage is influenced by $G_\text{K1}$ and pump currents.\\
\midrule

\multicolumn{3}{l}{\citet{del2020sensitivity}  } \\
\blt To identify which of the uncertain inputs mostly affect electrophysiology of the left ventricle.
& \blt Variance based sensitivity analysis. \blt Cardiac cell model: \citet{ten2006alternans}. 
& \blt The most influential input parameters for AP duration and shape similarity are the maximal IKr, IKs and ICal conductances ($G_\text{Kr}$, $G_\text{Ks}$ and $G_\text{CaL}$) along with the extracellular $Ca$ and $Na$ concentration.\\
\midrule

\multicolumn{3}{l}{\citet{chang2015uncertainty}  } \\
 \blt To investigate the impact of variability in ionic currents on the electrophysiological properties of human ventricular cells.
        & \blt Variance based sensitivity analysis. \blt Cardiac cell model: \citet{courtemanche1998ionic}. & \blt AP upstroke (max dV/dt) and maximum voltage were
        highly sensitive to $G_\text{Na}$. Dome voltage was most sensitive to
        $G_\text{Kur}$ and $G_\text{CaL}$, whilst $G_\text{K1}$ had a strong effect on $A_\text{90}$ and resting voltage. $A_\text{50}$ was most sensitive to $G_\text{CaL}$.\\
\midrule

 \multicolumn{3}{l}{\citet{pathmanathan2019comprehensive}  } \\
         \blt To demonstrate feasibility of performing comprehensive UQ/SA for cardiac cell models and demonstrate how to assess robustness and overcome model failure when performing cardiac UQ analyses.
        & \blt Variance based sensitivity analysis. \blt Human ventricular model: \citet{ten2006alternans}. & \blt $E_\text{h}$, plays a dominant role in determining MaxUpstrokeVelocity. TimeOfMaxUpstrokeVelocity is
       controlled by three parameters $E_\text{m}$, $k_\text{m}$ and $E_\text{z}$. ($E_\text{h}$, log($\delta_\text{h}$), $E_\text{r}$, $E_\text{d}$, $E_\text{f}$), which were the parameters identified to be highly influential.\\
\midrule

\multicolumn{3}{l}{\citet{mora2017sensitivity}  } \\
\blt To study the electrical activity and ionic homeostasis of failing myocytes. \blt To compare with experimental data.
& \blt Single-parameter sensitivity analysis. \blt Human ventricular model: modified \citet{o2011simulation}. 
& \blt $I_\text{NaL}$ and $I_\text{NaK}$ are the most important contributors to AP duration variations. SERCA plays an important role in modulating $Ca^{2+}$, with the $Na^{2+}$/$Ca^{2+}$ exchanger (NCX) and other $Ca^{2+}$ cycling proteins also playing a significant role.\\
\midrule

\multicolumn{3}{l}{\citet{sadrieh2013quantifying}  } \\
\blt To extend existing sensitivity analyses to electrocardiogram (ECG) signals derived from multicellular systems and quantify the contribution of ionic conductances to emergent properties of the ECG.
& \blt Partial least squares analysis. \blt Human ventricular model: \citet{o2011simulation}. 
& \blt  $A_\text{90}$ is approximately fourfold more sensitive to changes in $G_\text{CaL}$ than in isolated cells but less sensitive to changes in $G_\text{K1}$ and $G_\text{ncx}$.\\
\midrule

\multicolumn{3}{l}{\black{\citet{Parikh2019}}  } \\
\blt \black{To apply global sensitivity analysis to 
the existing Comprehensive in vitro Proarrhythmia 
Assay (CiPA) in silico framework to identify the key model 
components that derived metrics are most sensitive to}.
& \blt \black{Variance based sensitivity
analysis.} \blt \black{Human ventricular model: CiPAORd model, modified from \citet{o2011simulation}}. 
& \blt  \black{The Sobol 
sensitivity indices indicate that $A_\text{90}$ is the most sensitive to sbIKr block, qNet to sbINaL, and peakCa to bICaL.}\\

\end{longtable}

% Literature review of previous work
There is a substantial body of literature on sensitivity analysis of
detailed cardiac AP models, with a selection of the
most notable studies summarized in Table \ref{tab:literature}. Among
these, only the work of \cite{romero2011systematic} specifically
addressed rabbit ventricular AP models. Their study
systematically investigated how parameter variations affect a wide
range of electrophysiological properties, including steady-state AP
and intracellular calcium concentrations ($[\text{Ca}^{2+}]_i$), AP
duration and triangulation, AP duration rate dependence and restitution
curves, and AP duration adaptation to abrupt changes in basic cycle length
(BCL).
A major limitation of \cite{romero2011systematic} is its use of local
sensitivity analysis, where parameters are varied by fixed percentages
($\pm15$\% and $\pm30$\%) around Shannon model baseline values. This
approach is valid only near reference values and ignores interactions
between multiple parameters. Experimental data from
\citep{Lachaud2022} show significant variability that the baseline
Shannon model cannot capture, and its nonlinearity suggests parameter
sensitivity varies across the space. Thus, global sensitivity
analysis, which evaluates parameter influence across the full space
and considers nonlinear interactions, is essential for a more complete
understanding of model behaviour. 

% Global sensitivity analysis
Several methods for global sensitivity analysis, including partial
rank correlations and Fourier amplitude sensitivity, are
well-documented in the literature
\citep{Renardy2019}.
For this study, we have chosen the Sobol global sensitivity method
\cite{sobol2001global},
based on its robustness and efficiency.
Sobol's method quantifies the contribution of input variables or
parameters to the variance of a model's output, helping to identify
the most influential factors. A brief informal outline of the
theoretical concepts behind Sobol analysis is provided in further below.

% Specific objectives
Our work has three main objectives. First, we estimate Sobol
sensitivity indices for all maximal current strength parameters in the
Shannon model, focusing on their influence on selected AP biomarkers,
including duration, plateau potential, and resting potential. Second,
we rank these parameters based on their grand total Sobol sensitivity
indices to identify the most influential ionic currents. Third, we
validate these rankings and propose a hierarchy of reduced models,
demonstrating that dimensionality reduction is feasible without
significant loss of predictive accuracy.

\section{Methods and models}
\label{section:model}

\subsection{Shannon's AP model and output biomarkers}
\label{section:AP model}

% AP model equations
To understand the cell-to-cell variability reported in
\citep{Lachaud2022}, we consider the
\citet{shannon2004mathematical} mathematical model for the rabbit
ventricular myocyte AP. This model, along with that of
\citet{Mahajan2008}, is among the two widely used models for this cell type.
We chose the Shannon model as it better captures the measurements of
\citet{Lachaud2022}, as detailed in their analysis. 
% The Shannon model incorporates, in particular, a subsarcolemmal
% compartment, together with junctional and bulk sarcolemmal
% compartments, physiologically realistic cytosolic calcium buffering
% parameter values, a reversible sarcoplasmic reticulum calcium pump,
% a sodium-calcium exchanger that is dependent on the concentration of
% intracellular sodium and is regulated by calcium, and also a model
% of sarcoplasmic reticulum calcium release. 
As typical for models of myocyte electrophysiology, the Shannon model
takes the form of a system of ordinary differential equations
\begin{equation} \label{Shannon}
\begin{aligned}
 &\frac{d}{d t} V(t) =-\left(\sum_{i=1}^{N} p_i I_i(V, \mathbf{z}, \rev{\mathbf{c}}, \boldsymbol{\theta})+I_{\text {stim }}(t,\mathbf{u})\right), \\
 &\frac{d}{d t} \mathbf{z}(t)=\mathbf{g}(V, \mathbf{z}, \boldsymbol{\theta}) , \\
 &\rev{\frac{d}{d t} c_k(t)=s_k\left(\sum_{j_k} p_{j_k} I_{j_k} (V, \mathbf{z}, \mathbf{c}, \boldsymbol{\theta})\right), ~~~~~ j_k \subseteq \{ x \in \mathbb{N} \mid 1 \leq x \leq N\}}.
\end{aligned}   
\end{equation}
Here, $V$ is the electric potential across the cell membrane, $t$
represents time. \rev{Vector $\mathbf{z}$
represents variables describing 
channel gating configurations such as activation and inactivation or
Markov model states, and vecor $\mathbf{c}$, with components $c_k$,
represents ionic intracellular concentrations, while $\mathbf{g}$ and $s_k$ are functional dependences.} 
Model parameters, such as maximal conductances and channel kinetics,
are contained in the vector $\boldsymbol{\theta}$, while $\mathbf{u}$
represents external protocol parameters, including stimulus timing,
duration, and strength. 
The transient changes in membrane potential, known as APs, are controlled by the sum of currents $I_i$
flowing across the membrane or between internal compartments (e.g.,
organelles). The Shannon model includes $N = 15$ distinct
currents, detailed in Table \ref{tab:des}. Their explicit formulations
and parameter values are available in \citep{shannon2004mathematical}. 
\black{\rev{The factors $p_i$ multiply the ionic currents everywhere
    the latter appear}, and  represent increase or decrease of current
    strengths relative to their baseline values. They have been
    embedded in the model for the purposes of the sensitivity
    analysis, as they allow the original model formulation to be used
    without modification.}  
% AP initial conditions
A stimulus current, $I_{\text{stim}}$, is used to excite the cell and
is applied at a basic cycle length (\( t_{\text{BCL}} \)). Under
physiological pacing conditions, the system reaches a periodic train
of APs irrespective of the initial conditions. %For this study, \black{the stimulus amplitude was set to 9.5 A/F} and trains
%of 1000 APs were generated with \( t_\text{BCL} = 500 \, \mathrm{ms}
%\). The final AP from each train was used to perform the sensitivity
%analysis. 

\begin{table}[t]
\begin{center}
\begin{tabular}{ll}
\toprule
Ionic current, $I_i$ & Description   \\
\midrule
 $I_\text{CaL   }$   & L-type Ca$^{2+}$ current.             \\
 $I_\text{Cab    }$   & Background Ca$^{2+}$ current.              \\
 $I_\text{Cap    }$   & Ca$^{2+}$ pump current. \\
 $I_\text{ClCa    }$   & Ca$^{2+}$-activated Cl$^{-}$ current.               \\
 $I_\text{K1   }$   & Inward rectifier K$^{+}$ current / Time-independent K$^{+}$ current.                   \\
 $I_\text{Kp    }$   & Background potassium K$^{+}$ current.                \\
 $I_\text{Kr   }$   & Fast delayed rectifier K$^{+}$ current.                 \\
 $I_\text{Ks     }$   & Slow delayed rectifier K$^{+}$ current.                \\
 $I_\text{Clb    }$   & Cl$^{-}$ background current.                   \\
 $I_\text{Na    }$   & Fast Na current.              \\
 $I_\text{NaCa }$   & Na$^{+}$/Ca$^{2+}$ exchanger current.                \\
 $I_\text{NaK     }$   & Na-K pump current.                \\
 $I_\text{Nab      }$   & Background Na$^{+}$ current.                     \\
 $I_\text{tof   }$   & Fast component of the transient outward potassium current.               \\
 $I_\text{tos    }$   & Slow component of the transient outward potassium current. \\
\bottomrule
\end{tabular}
\caption{The fifteen ionic currents of the \citet{shannon2004mathematical}
 mathematical model for the rabbit
ventricular myocyte AP. \label{tab:des}}
\end{center}
\end{table}

The model equations are solved numerically. To avoid coding errors,
we use a machine readable model specification file available from
the CellML model repository \citep{Lloyd2008}. Numerical integration is
done using the Myokit suite for cardiac cellular electrophysiology
simulations \citep{Clerx2016}, which in turn employs solvers for nonlinear 
differential/algebraic equation from the SUNDIALS library
\citep{Hindmarsh2005}. 

% Quantities included in the sensitivity analysis
% - parameters
Due to the large number of parameters in the Shannon model, it is not
computationally feasible to include all in a sensitivity
analysis. Following previous studies
e.g.~\citep{romero2011systematic,Lachaud2022}, we restrict the
attention to investigating the influence of the relative strengths of ionic
currents. To measure these we introduce $N=15$ auxiliary parameters
denoted by $p_i$ in equations \eqref{Shannon}. These can be interpreted as
scaling factors of the maximal conductances for currents of the Ohmic
form, or of the $\max$ $I_i$ for currents of the
Goldman-Hodgkin-Katz form.  In
essence, the factors \( p_i \) represent the relative strengths of
ionic currents compared to their ``baseline'' values published in
\citep{shannon2004mathematical}.

% -output biomarkers
The influence of the sensitivity parameters $p_i$ on the model will be
assessed by measuring the variation of $K=6$ biomarkers \black{$y_i$} that capture
important characteristics of the AP waveform $V(t; p_1,p_2,\ldots,
p_N)$. These model output quantities 
are listed and defined in Table \ref{tab:AP feature}. 
In the Table and throughout the rest of the text $t_0$ and $t_\text{BCL}$
refer to the start and the end of the last AP, and $V(t)$ refers to its
voltage trace. 
%\black{To ensure that biomarker values are comparable in magnitude as required by the Sobol analysis detailed in the next section, these have been normalized by their standard z-score function $y_i = (b_i - \mu_i)/\sigma_i$, where $\mu_i$ is the mean of $b_i$ and $\sigma_i$ is its standard deviation.}
In summary, the relationships between biomarkers and
parameters can be represented formally by a mapping $\mathbf{f}$ as
\begin{gather}
  \label{eq:model}
  \mathbf{y} = \mathbf{f} (\mathbf{p}), ~~~~~\mathbf{p} = (p_1,p_2,\ldots,p_N).
\end{gather}

\begin{table}[t]
  \begin{center}
      \begin{tabular}{p{0.29\textwidth}p{0.09\textwidth}p{0.51\textwidth}}
    \toprule
    AP biomarker & Symb $y^k$ & Definition (Description)   \\
    \midrule
  Peak potential
  & $V_\text{max}$
  & $V_\text{max} = \max_t{V(t)}$. % The maximum voltage value of the
                                % membrane potential following
                                % stimulus 
  \\
 Resting potential \black{(alt. maximal diastolic potential)}
 & $V_\text{rest}$
 & $V_\text{rest} = V(t_\text{BCL})$,

 where $t_{\text{BCL}}$ is the
 ``basic cycle length''. \\ % The membrane potential immediately  before the next stimulus.
    AP duration at 90\% repolarization
    & $A_{90}$
    & $A_{90} = V^{-1}\big(V_\text{max} - 0.9 |V_\text{max} -
    V_\text{rest}|\big) - t_0$
    
   (The time from stimulus to the 
  moment where the potential reaches 90\% of full
  repolarization as defined by the difference between peak and resting potential.) \\
  AP duration at 30\% repol.
&   $A_{30}$
& Ditto. \\
 \black{Plateau potential} 
 & \black{$V_\text{plt}$}
 &  \black{$V_\text{plt}  = V(t_0 + {A_{90}}/{2})$}. \\ %  Voltage at the plateau phase of the AP.
 Bulk
 & $B$ 
 & $B = \int_{0}^{t_{\text{BCL}}} |V(t)-V_{\text{rest}}| \,dt. $ \\
% (measures the ``bulk'' of the AP).\\
  \bottomrule
  \end{tabular}
  \caption{A list of the six AP output biomarkers $y^k$ used for
    sensitivity analysis, and definitions of their corresponding
    relationships to the parameters of the Shannon model, $y^k =
    f^k(p_1,p_2,\ldots, p_N)$. The dependence of $y^k$ on parameters
    is due to  the voltage potential being parameter dependent, $V =
    V(t;  p_1,p_2,\ldots, p_N)$, as discussed in the text.
    \label{tab:AP feature}}
  \end{center}
\end{table}

\subsection{Sobol's sensitivity analysis of variance decomposition}
\label{sec:sobol}

We employ a global method for sensitivity analysis based on variance
decomposition, originally proposed by \cite{sobol1976uniformly}. The starting point is
the relationship \eqref{eq:model} between AP output biomarkers \(
\mathbf{y} \) and the relative strengths of ionic currents \(
\mathbf{p} \). The sensitivity of each biomarker \( y^k \) can be
analyzed independently of the others. For simplicity, we drop the
superscript \( k \) and focus on a single scalar model output, \( y =
f(\mathbf{p}) \). \black{To ensure that biomarker values are comparable in magnitude, the values are then normalized
$\tilde{y} = \tilde{f}(\mathbf{p})$
by computing their standard z-score $\tilde{f}(\mathbf{p}) = z(f(\mathbf{p}))$, defined as $z(x_i) = (x_i - \mu)/\sigma$ for a general set of values $\{x_i, ~ i = 1,\dots, L\}$ with mean and standard deviation $\mu$ and $\sigma$, respectively.}

% Functional decomposition
Consider the parameter vector as a vector of \( N = 15 \) independent random variables, \( \mathbf{P} = (p_1, p_2, \ldots, p_N) \). The \black{standardized} biomarker \(\tilde{ Y} \) is then also a random variable, given by \( \tilde{Y} = \tilde{f}(\mathbf{P}) = f(p_1, p_2, \ldots, p_N) \), where \( f \) is a deterministic function specified in Table \ref{tab:AP feature} and consequently so is $\tilde{f}$. It can be shown that any function of independent random variables has a functional decomposition of the form
\begin{subequations}
\label{eq:func_decomp}
\begin{gather}
  \label{eq:func_decomp_a}
  \tilde{f}(\mathbf{P}) = \tilde{f}_0 + \sum_{i=1}^{N} \tilde{f}_i(p_i) + \sum_{1 \leq i < j \leq N} \tilde{f}_{ij}(p_i, p_j) + \cdots +
  \tilde{f}_{1,2,\ldots,N}(p_1, p_2, \ldots, p_N),
\end{gather}
where the quantities \( \tilde{f}_I(p_I) \) are defined as:
\begin{gather}
  \label{eq:func_decomp_b}
  \tilde{f}_I(p_I) = \mathbb{E}[\tilde{f}(\mathbf{P}) \mid p_I] - \sum_{J \subset I} \tilde{f}_J(p_J),
\end{gather}
\end{subequations}
for any subset of indices \( I \subseteq \{1, 2, \ldots, N\} \), and \( \mathbb{E}[\cdot] \) is the expectation value operator. These quantities represent the interaction “effects” on the output \( \tilde{Y} \) of all variables indexed by the set \( I \), where the sum is taken over all proper subsets \( J \) of \( I \). For example,
\begin{gather*}
\tilde{f}_0 = \mathbb{E}[\tilde{f}(\mathbf{P})], \\
\tilde{f}_i(p_i) = \mathbb{E}[\tilde{f}(\mathbf{P}) \mid p_i] - \tilde{f}_0, \\
\tilde{f}_{ij}(p_i, p_j) = \mathbb{E}[\tilde{f}(\mathbf{P}) \mid p_i, p_j] - \tilde{f}_i(p_i) - \tilde{f}_j(p_j) - \tilde{f}_0,
\end{gather*}
represent the overall mean value of the output, the correction effects due to the isolated “action” of each single parameter (first-order effects), and the correction effects due to the interactions of all possible combinations of two parameters (second-order interaction effects), respectively.
The decomposition \eqref{eq:func_decomp_a} can be easily verified by
nested back-substituting \eqref{eq:func_decomp_b} into
\eqref{eq:func_decomp_a}, and taking into account the fact that the conditional
expectation of a deterministic function in the case all of its
values are deterministic is equal to the value of the function itself, formally
\[
\mathbb{E}[\tilde{f}(p_1, p_2, \ldots, p_N) \mid p_1 = p_1, p_2 = p_2, \ldots,
  p_N = p_N] = \tilde{f}(p_1, p_2, \ldots p_N).
\]

% Variance decomposition
The standard measure of the variation of the biomarker \( \tilde{Y} \) is the variance \( \mathbb{V}[\tilde{f}(\mathbf{P})] \). Substituting the functional decomposition \eqref{eq:func_decomp} into the definition of variance:
\[
\mathbb{V}[\tilde{f}(\mathbf{P})] = \mathbb{E}[(\tilde{f}(\mathbf{P}) - \mathbb{E}[\tilde{f}(\mathbf{P})])^2],
\]
expanding the products, and using the fact that cross terms are orthogonal
\[
\mathbb{E}[\tilde{f}_I(P_I)\tilde{f}_J(P_J)] = 0 \quad \text{for} \quad I \neq J, ~~~~ I, J \subseteq \{1, 2, \ldots, N\},
\]
along with the property
\[
\mathbb{V}[\tilde{f}_I (P_I)] = \mathbb{E}[\tilde{f}_I (P_I)^2],
\]
we find that the total variance has the decomposition:
\begin{gather}
  \label{eq:var_decomp}
\mathbb{V}[\tilde{f}(\mathbf{P})] = \sum_{i=1}^{N} \mathbb{V}[\tilde{f}_i(p_i)] + \sum_{1 \leq i < j \leq N} \mathbb{V}[\tilde{f}_{ij}(p_i, p_j)] + \cdots + 
\mathbb{V}[\tilde{f}_{1,2,\ldots,N}(p_1, p_2, \ldots, p_N)].
\end{gather}

% Definitions of Sobol indices
With this, direct measures of the ``main effect'' of a given parameter
$p_i$, and the ``higher-order effects'' of interactions between
combinations of parameters on the output $\tilde{Y}$ are defined as the
corresponding ``partial'' variances in the decomposition \eqref{eq:var_decomp}
normalised by the total variance,
\begin{subequations}
  \label{eq:indices}
\begin{gather}
\label{eq:index}
S_i = \frac{\mathbb{V}[\tilde{f}_i(p_i)]}{\mathbb{V}[\tilde{f}(\mathbf{P})]}, ~~~
S_{ij} = \frac{\mathbb{V}[\tilde{f}_{ij}(p_i,p_j)]}{\mathbb{V}[\tilde{f}(\mathbf{P})]}, ~~~
S_{I} = \frac{\mathbb{V}[\tilde{f}_{I}(p_I)]}{\mathbb{V}[\tilde{f}(\mathbf{P})]}.
\end{gather}
These quantities are known as first-order, second-order and higher-order
Sobol sensitivity indices, respectively, and represent fractional sensitivities
in the sense that they add up to unity,
$$
1 = \sum_{i=1}^N S_i + \sum_{1 \leq i < j \leq N} S_{ij} + \cdots + S_{1,2,\ldots,N}.
$$
Finally, the ``total effect'' of a parameter $p_i$ is quantified by
the so called Sobol total-effect index, defined as the sum of all
sensitivity indices related to this parameter and its possible interactions,
\begin{gather}
\label{eq:total}
S_{T_i} = S_i + \sum_{j \neq i} S_{ij} + \sum_{\substack{j \neq i, k \neq i \\ j < k}} S_{ijk} + \cdots + S_{1, 2, \ldots, N}.
\end{gather}

% Estimates for Sobol indices by Monte-Carlo random sampling
\black{The Sobol sensitivity indices \eqref{eq:indices} are estimated via
quasi-random sampling. To achieve this, the parameter space is sampled using
a quasi-Monte Carlo method, which generates the
so-called ``Sobol sequences'' -- sequences of low-discrepancy
quasi-random numbers designed to fill space uniformly
\citep{sobol1976uniformly}.} The values of the function \( \tilde{f} \) and the
terms of its functional decomposition \eqref{eq:func_decomp} are
evaluated at these sequences. Finally, the ``Jansen'' estimator
\citep{jansen1999analysis} is used to calculate
the Sobol sensitivity indices from the sampled data.  
For this study, we employ an implementation of this procedure
available from the SALib Sensitivity Analysis Library in Python
\citep{Herman2017}. Further technical details are omitted
here, except to note that the consistency and bias of index estimation
depend on the number \( M \) of randomly sampled points in the
parameter space. This dependence is investigated further below.

% Confidence interval of the Sobol index estimates (bootstrap sampling distributions)
Sobol index estimates are random values drawn from a sampling
distribution. To assess their uncertainty, bootstrapping is used
\citep{efron1993introduction}. The original sample is resampled with
replacement 1000 times, generating bootstrap samples from which Sobol
indices are recalculated. This process constructs empirical sampling
distributions, providing 95\% confidence intervals for the indices and
assessing variability due to finite sample size \( M \). Bootstrapping
is preferred over repeated Monte Carlo estimations, which are
computationally more expensive.

\subsection{Parameter ranking and dimensionality reduction}
\label{modelreduction}

Sensitivity studies lead to a rank of sensitivity parameters
in descending order of their influence on the model output. The
ranking can be further exploited to reduce the dimensionality of the
parameter space of the problem.

% Ranking
A number of different criteria for parameter ranking may be
employed. In this study we define a ``grand'' total Sobol sensitivity  
index for each of the sensitivity parameters \black{$p_i$}, as
\begin{gather}
\label{eq:grand_total}
S_{Gi} = \sum_{k=1}^K w_k S^k_{Ti}.
\end{gather}
\end{subequations}
Here $S^k_{Ti}$ is the total Sobol index given by equation
\eqref{eq:total} which measures the overall influence of
the parameter $p_i$ on a single AP biomarker $y^k$. In contrast the grand
total index $S_{Gi}$ measures \black{the} overall influence of
the parameter $p_i$ on all AP biomarkers $\mathbf{y}$. The weights
$w_k$ can be used to tune the importance of the outputs from any
a priori considerations; here we take $w_k = 1, k=1, \ldots, K$ \black{as we consider all biomarkers to be of equal interest}. 
%\black{Although the AP biomarkers, especially bulk, vary a lot in magnitude, the results of the Sobol analysis do not change as biomarkers have been standardized as mentioned above}.
The sensitivity
parameters can then be ranked in descending influence by sorting the
values of their grand total Sobol sensitivity indices. In the
following, we denote parameter vectors where the components are
ordered in decreasing grand total Sobol index by angular brackets
\begin{gather}
\label{eq:ordered}
 \langle\mathbf{p}\rangle = \langle p_1, p_2, \ldots, p_N \rangle, ~~~~~~~  S_{G1} \ge S_{G2} \ge
\ldots \ge S_{GN},
\end{gather}
and distinguish it from parameter vectors $\mathbf{p}$ where the order
of components is immaterial. 

% reduction of dimensionality of parameter space
If parameters are independent, the
dimensionality of the parameter space can be reduced by keeping the
values of a subset of $N-M$ parameters $(M \in[0,N])$ fixed to ``hard-coded''
constants. The natural choice is to fix the parameters with smallest grand total
Sobol sensitivity indices as their influences on the model output are
relatively less significant. Whether this is an acceptable
approximation can be measured at any fixed point in the
parameter space by the relative difference/error
\begin{subequations}
\begin{gather}
e^{\langle M \rangle}= \frac{1}{K} \sum_{k = 1}^K \left|\frac{(y_j^{\langle M \rangle,k} - y_j^k)}{y_j^k}\right|,
\end{gather}
between the output vectors of the reduced and the ``full'' models given
by 
\begin{align}
  \mathbf{y}^{\langle M \rangle} &= \mathbf{f}(\langle p_1,p_2,\ldots  p_M,1,\ldots,1 \rangle),  \label{reduce} \\
  \mathbf{y}&= \mathbf{f}(\langle p_1,p_2,\ldots  p_M,p_{M+1},\ldots, p_N \rangle),  \label{full} 
\end{align}
\end{subequations}
respectively.

To obtain a global measure of the discrepancy over the entire
parameter space we use quasi random Monte-Carlo sampling (similarly to
assessing variability in section \ref{sec:sobol}), and compute the mean relative
error from the generated samples of parameter vector values 
\begin{gather}
\label{eq:error}
E^{\langle M \rangle}= \frac{1}{LK} \sum_{j = 1}^L \sum_{k = 1}^K
\left|\frac{(y_j^{\langle M \rangle,k} - y_j^k)}{y_j^k}\right|,
\end{gather}
where $L=2000$ is the number of samples.
This has the advantage over a mean squared error or over a Euclidian
norm that it can be interpreted directly as the relative 
average discrepancy between full and reduced models.
When $E = 0$ agreement is perfect.
Using the relative error has the further advantage of
non-dimensionalising the errors of individual
components which are otherwise measured in different units and may have
significantly different magnitudes.

\section{Results and discussion: Sensitivity analysis of the Shannon model}
\label{section:RandD}
\begin{figure}[t]
\centering
\includegraphics[width=0.9\textwidth]{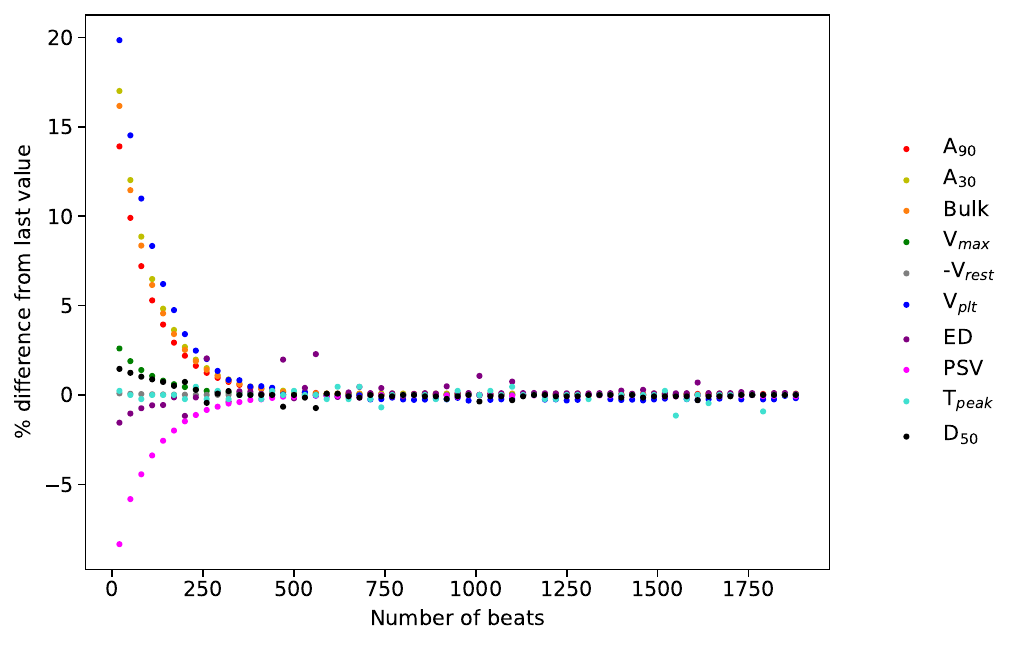}
   \caption{\rev{Biomarkers as a function of the length of AP
       train. The percentage differences in biomarker values from
       their respective values measured in the 2000th AP are plotted
       for increasing number of pre-pacing beats. The simulations are
       performed at 50\% block of \( I_\text{NaK} \) and 70\% block of
       \( I_\text{NaCa} \) and all other parameters at baseline
       values. Biomarkers are specified in the figure legend.} \label{fig000}}  
\end{figure}

\black{In the following we present and discuss the results of applying the Sobol sensitivity analysis for a standard setup where the the stimulus amplitude was set to 9.5 A/F and trains
of 1000 APs were generated with \( t_\text{BCL} = 500 \, \mathrm{ms}
\). Biomarkers were measured in the final AP from each train. Sample
sequences of such measurements with size $M=8192$ were used to then
evaluate the sensitivity indices. We investigated the effects of
varying these standard setup choices, and found that the results and
conclusions do not differ significantly. As an illustration, \rev{the
percentage differences in biomarker values from their respective values
measured at the 2000th beat are plotted} in Fig~\ref{fig000} as a function
of the length of the train, and show that trains of 1000 APs are
sufficiently long to reach the steady state in the simulations \rev{even
  though rare random fluctuations smaller than 2\% may occur
  occasionally due to accumulation of numerical error. } 
The other standard setup choices are also justified further below.
}

\subsection{The parameter range of normal response}
\label{sec:subspace}
% Instabilities across the parameter space
To fulfill their physiological functions, biological cells exhibit a
variety of complex behaviors under different conditions. Similarly,
mathematical models of the AP, such as the Shannon model
\eqref{Shannon}, have qualitatively distinct solutions across
their parameter space due to their non-linear nature \citep{Qu2014}.  The
simplest response to periodic stimulation \( I_\text{stim}(t)
\) is a rapid return to a stable equilibrium (resting state),
known as a 1:0 response, which mimics non-excitable cells. At other
parameter values, a bifurcation to a periodic solution occurs, where
each stimulus elicits a single AP, producing a 1:1 response that
models normal physiological behaviour. Secondary bifurcations from
the 1:1 response can result in a 2:1 response, where one stimulus
generates an AP while the next does not, or a 2:2 response, where
alternating stimuli produce long and short APs in a stable 2-periodic
pattern. This 2:2 response, known as alternans, is thought to reflect
early signs of instability. Further tertiary instabilities can lead to
chaotic regimes, representing abnormal electrophysiological behavior,
which may culminate in fibrillation. \black{The review article of \citep{Qu2014} discusses nonlinear and stochastic dynamics in the heart}.

% Restriction to and construction of the parameter subspace of normal response
Naturally, sensitivity to parameters is expected to vary significantly
across different solution regimes. We restrict our sensitivity analysis
of the Shannon model to the parameter region corresponding to the
normal 1:1-response, as this represents the most common physiological
behavior.  
To perform this analysis, it is first necessary to identify the region
in the parameter space where the primary normal response occurs. This
region, occasionally called a ``Busse balloon'' in the context of
pattern-forming dynamical systems \citep{Cross1993}, is not 
the primary focus of this study. Instead, our goal is to identify a
sufficiently large region of normal response suitable for sensitivity
analysis.  To achieve this, we begin with the baseline Shannon model,
where \( \{p_i = 1\}_{i=1}^N \), as the baseline model is calibrated
to produce a normal response. Each sensitivity parameter \( p_i \) is
then varied independently within the range \( 10^{-4} \) to \( 10 \),
until the first transition to a secondary instability or a
non-excitable state is observed.

\begin{figure}[t]
\centering
\includegraphics[width=0.95\textwidth]{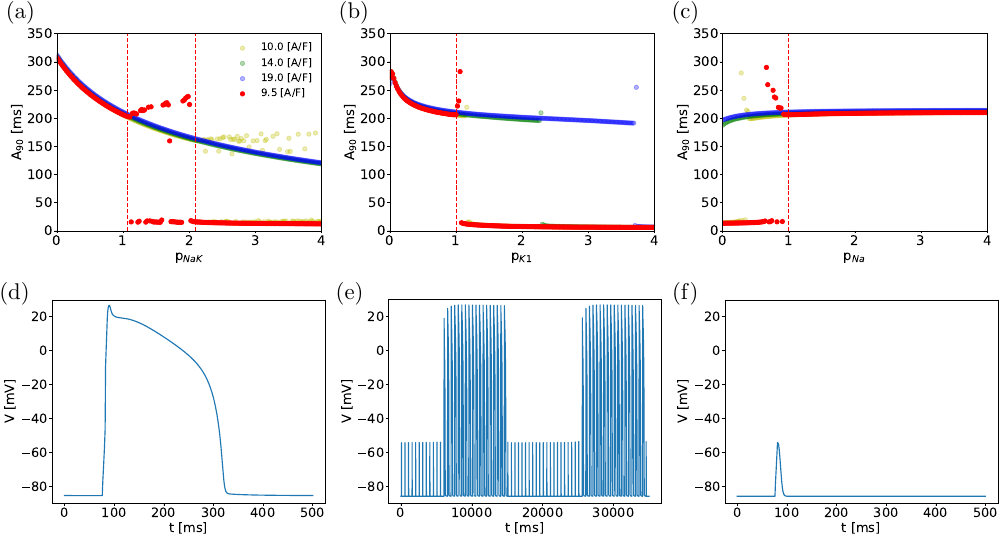}
\caption{
(Top row) Ranges of parameter values for normal response illustrated on a selection of parameters \black{and for various stimulus amplitudes}.  
\black{Values of $A_{90}$ as functions of $p_{\text{NaK}}$,  $p_{\text{K1}}$ and $p_{\text{Na}}$ in (a), (b) and (c), respectively, and for four stimulus amplitudes as listed in the legend. The
red dashed lines outline regime boundaries (for amplitude 9.5 A/F) as discussed in the text.}
(Bottom row) Types of response to stimuli observed for various values of the
maximal density of the sodium-potassium pump current $p_{\text{NaK}}$ for amplitude 9.5 A/F.
\black{(d)}:
Normal response where each stimulus elicits a single 1:1 AP is
observed for $p_{\text{NaK}} < 1.07$;   (e):  An "intermittent"
excitation pattern where $Q$ successive stimuli elicit $Q$ normal
APs, while the next $Q$ stimuli do not, observed for
$p_{\text{NaK}}\in(1.07,2.1)$  (f): Stimuli do not elicit an AP
response for $p_{\text{NaK}} > 2.1$.} 
    \label{fig010}
\end{figure}

Fig \ref{fig010} illustrates the procedure for the parameter 
\( p_\text{NaK} \). One biomarker, the AP duration at 90\% repolarization (\(
A_{90} \)), is used to assess the model's response, as other
biomarkers exhibit similar behavior. At \( p_\text{NaK} = 1 \), a
normal 1:1-response is recorded, as shown in Fig
\ref{fig010}\black{(d)}. As a side note: Fig
\ref{fig010}\black{(d)} which also depicts a typical AP profile in this
model.  The normal response persists for \(
p_\text{NaK} \) in the range \([10^{-4}, 1.07]\). Beyond \(
p_\text{NaK} = 1.07 \), an intermittent response emerges, where
approximately 20 normal APs are followed by an equally long
equilibrium phase, as illustrated in Fig \ref{fig010}\black{(e)}. This
behavior produces a two-valued curve in the \( (p_\text{NaK}, A_{90})
\) plane, with \( 
A_{90} \approx 0 \) during equilibrium and finite values during the normal
response sequence. For \( p_\text{NaK} > 2.1 \), only the equilibrium
response is observed, as shown in Fig \ref{fig010}\black{(f)}.
%\black{Supplementary Figure
%XXX ref{fig:range}} includes results for all other parameters, while
Table \ref{tab:range} summarizes these findings and defines the parameter
subspace for the normal response region analyzed in this study. 

\black{The threshold of excitation at baseline parameter values is approximately 9.4 A/F, so our standard setup uses a slightly supercritical stimulus amplitude. Increasing the amplitude of the stimulus current enlarges the region of normal response as shown in Fig \ref{fig010}(a,b,c) where results for amplitudes up to 2 times the threshold are also included. However, increasing the stimulus amplitude does not significantly alter the sensitivity ranking of parameters as will be shown further below.} 

\begin{table}[t]
\begin{center}
\begin{tabular}{ll|ll|ll}
\toprule
Parameter & Range   & Parameter & Range & Parameter & Range\\
\midrule
 $p_{\text{NaK}}$   & {[}0.0001, 1.07{]}             &
 $p_{\text{Kr}}$     & {[}0.0001, 3.2{]}             &
 $p_{\text{tos}}$    & {[}0.0001, 2.5{]}             \\
$p_{\text{K1}}$    & {[}0.03, 1.01{]}                & 
$p_{\text{NaCa}}$   & {[}0.6, 3{]}                   &
 $p_{\text{Clb}}$    & {[}0.0001, 10{]}              \\
 $p_{\text{CaL}}$    & {[}0.7, 1.6{]}                 &
 $p_{\text{Ks}}$     & {[}0.0001, 10{]}              &
 $p_{\text{Cab}}$    & {[}0.34, 10{]}                 \\
$p_{\text{Cap}}$    & {[}0.0001, 4.5{]}               &
 $p_{\text{ClCa}}$ & {[}0.0001, 10{]}                 &
 $p_{\text{Kp}}$     & {[}0.0001, 10{]}               \\
 $p_{\text{Na}}$     & {[}1, 10{]}                    &
$p_{\text{Nab}}$   & {[}0.0001, 10{]}                 &
$p_{\text{tof}}$    & {[}0.0001, 4{]}                 \\
\bottomrule
\end{tabular}
\caption{Parameter subspace of normal response. Sensitivity analysis is performed within this range of parameter values. \label{tab:range}}
\end{center}
\end{table}

%It is shown in \cite{gemmell2014population} that there appeared to be
%no limitations  on the values of $p_{\text{Ks}}$. 
%IKp is the unique one that almost exert no influence on AP
%trace. Another interesting range is for INa starting from 1 and this
%may because the depolarization, also called the rising phase, is
%caused when positively charged sodium ions (Na$^{+}$) suddenly rush
%through open voltage-gated sodium channels so Na$^{+}$ must be large
%enough to activite the cell. We will study what happens for AP trace
%inside these ranges in the next section and use these refined ranges
%for 15 parameters for sensitivity analysis. 

\subsection{Consistency and bias of Sobol's index estimators}
\label{section:com}

% goal of section
The Sobol sensitivity indices, as statistical estimators derived from
random samples (Section \ref{sec:sobol}), must exhibit key properties
of good estimators, such as consistency (convergence to true values
with larger sample sizes) and unbiasedness (no systematic deviation
from true values). Here, we evaluate
these properties to ensure the robustness of our Sobol sensitivity
indices estimation. 

\begin{figure}[t]
\centering
\includegraphics[width=0.95\textwidth]{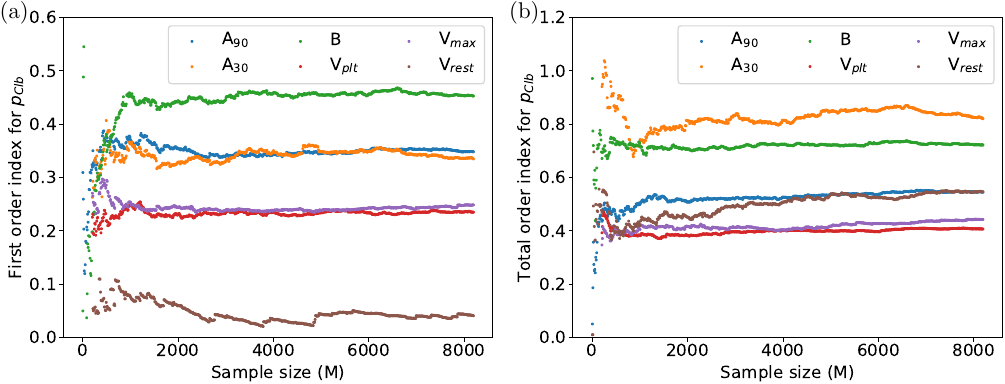}
  \caption{Consistency of Sobol sensitivity index estimation. Values of
  the first-order (a) and total-order (b) sensitivity indices
  of the background chloride current maximal density $p_\text{Clb}$
  for the six AP biomarkers of interest (as stated in the legend) when
  calculated with different numbers of
  samples.\label{fig020}} 
\end{figure}

% consistency
Fig \ref{fig020} demonstrates the  consistency of first-order and
total-order indices of the parameter $p_{\text{Clb}}$ in the
large-sample limit across the six AP biomarkers. Specifically, smaller
sample sizes $M$ show greater variability, particularly in the
total-order indices which are sensitive to interaction effects. As the
sample size $M$ is increased, both first-order and total-order indices
converge to specific values. The sensitivity indices of all other
parameters behave similarly. Based on this consistency test we have
fixed the sample size to $M=8192$ in our subsequent analysis.

\begin{figure}[t]
%  \vspace*{8mm}
%  \hspace*{5mm}
\centering
\includegraphics[width=0.95\textwidth]{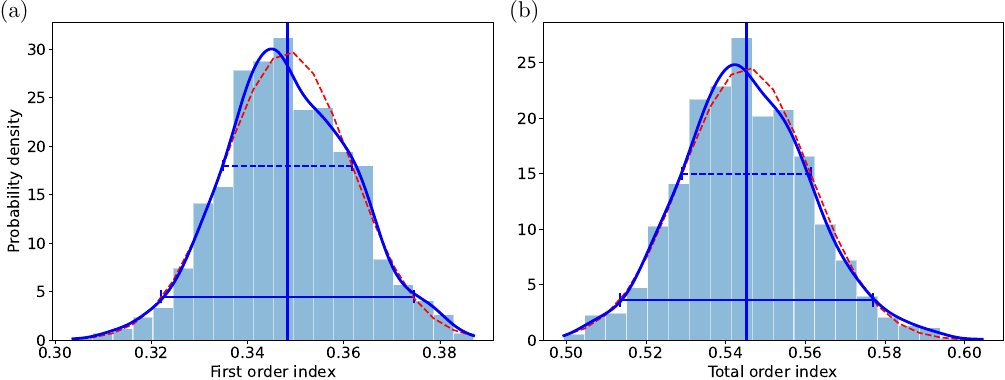}
 \caption{Bias of  Sobol sensitivity index estimation. Histograms of
   the bootstrapping distributions of the first-order (a) and
   total-order (b) sensitivity indices of \black{the} background
   chloride current maximal density $p_\text{Clb}$ for $A_{90}$
   constricted with a sample size of $M=8192$ are shown in blue. The
   mean value and the standard deviation of the bootstrapping
   histograms are shown by the vertical blue line and the horizontal
   blue bar, respectively. The latter are used to plot a normal
   distribution (red dashed line) for comparison. The kernel
   density estimates (KDE) bootstrapping distributions are shown as
   blue lines. \label{fig030} Horizontal blue bars
   show 1 (broken line) and 2 (solid) standard deviations from mean of the bootstraping
   distributions.} 
\end{figure}

% bias
To assess the bias of the Sobol sensitivity indices, we compute
bootstrap distributions of the first-order and total-order indices,
revealing their variability and central tendency. Comparing the mean
of the bootstrap estimates to the original Sobol indices allows
detection of systematic deviations and bias if any. Fig
\ref{fig030} shows the bootstrap distributions for \(
p_{\text{Clb}} \) with the AP feature \( A_{90} \). Both first-order
and total-order indices exhibit normal distributions, confirming
unbiased estimates at \( M = 8192 \). Similar unbiased behavior is
observed for all other parameters. 


% confidence intervals
Bootstrap distributions, such as those in Fig \ref{fig030},
provide a basis for constructing confidence intervals for the Sobol
sensitivity indices. Using the empirical “68–95–99.7” rule,
approximately 95\% of estimates lie within two standard deviations of
the mean. Confidence interval lengths computed in this way are
presented in Tables \ref{tab:summarytable} and \ref{tab:2nd} for all
Sobol sensitivity index estimates. 
\begin{table}[t]
\begin{center}
\begin{tabular}{l|l l l|l l l}
\toprule
$y^k$ &parameter & $S^k_{i}$ & 95$\%$ CI & parameter & $S^k_{Ti}$ & 95$\%$ CI \\
\midrule
    & $p_\text{Clb}$ & 0.33 & 0.032 & $p_\text{Clb}$ & 0.82 &0.065\\
$A_{30}$        & $p_\text{NaCa}$ & 0.063 & 0.019& $p_\text{Kr}$ & 0.49 & 0.056\\
        & $p_\text{ClCa}$ & 0.036 & 0.012& $p_\text{NaCa}$ & 0.47 & 0.05\\
\hline
 & $p_\text{Clb}$ & 0.35 & 0.023 & $p_\text{Clb}$ & 0.54 &0.036\\
$A_{90}$ & $p_\text{Kr}$ & 0.14 & 0.019& $p_\text{Kr}$ & 0.34 & 0.023\\
& $p_\text{NaCa}$ & 0.14 & 0.016& $p_\text{NaCa}$ & 0.24 & 0.018\\
\hline
 & $p_\text{Clb}$ & 0.45 & 0.027 & $p_\text{Clb}$ & 0.72 &0.03\\
$B$ & $p_\text{NaCa}$ & 0.076 & 0.015& $p_\text{Kr}$ & 0.26 & 0.025\\
& $p_\text{CaL}$ & 0.031 & 0.011& $p_\text{NaCa}$ & 0.24 & 0.023\\
\hline 
 $V_\text{plt}$ & $p_\text{Clb}$ & 0.23 & 0.021 & $p_\text{Clb}$ & 0.41 &0.022\\
& $p_\text{K1}$ & 0.17 & 0.020& $p_\text{K1}$ & 0.35 & 0.025\\
& $p_\text{CaL}$ & 0.15 & 0.017& $p_\text{Kr}$ & 0.26 & 0.025\\
\hline
& $p_\text{CaL}$ & 0.32 & 0.024 & $p_\text{Clb}$ & 0.44 &0.030\\
$V_\text{max}$  & $p_\text{Clb}$ & 0.25 & 0.024& $p_\text{CaL}$ & 0.40 & 0.022\\
& $p_\text{tos}$ & 0.050 & 0.013& $p_\text{K1}$ & 0.14 & 0.0098\\
\hline
&  $p_\text{K1}$ & 0.23 & 0.026 & $p_\text{Clb}$ & 0.55 &0.054\\
$V_\text{rest}$ &  $p_\text{Kr}$ & 0.068 & 0.028& $p_\text{Kr}$ & 0.54 & 0.052\\
&  $p_\text{Clb}$ & 0.041 & 0.030& $p_\text{K1}$ & 0.47 & 0.042\\
\bottomrule
\end{tabular}
\caption{The top three first-order and total-order sensitivity indices with corresponding confidence intervals for the six AP biomarkers considered. $S^k_{i}$ is the first Sobol index given by equation
\eqref{eq:index} which measures the individual influence of
the parameter $p_i$ on a single AP biomarker $y^k$ and $S^k_{Ti}$ is the total Sobol index given by equation
\eqref{eq:total} which measures the overall influence of
the parameter $p_i$ on a single AP biomarker $y^k$.} \label{tab:summarytable}
\end{center}
\end{table}

\begin{table}[t]
\begin{center}
\begin{tabular}{llll|llll}
\toprule
 $y^k$ & Parameter pair &$S^k_{ij}$ & 95$\%$ CI 
 &
 $y^k$ & Parameter pair &$S^k_{ij}$ & 95$\%$ CI  \\
 \midrule
$A_{90}$ & ($p_\text{Kr}$,$p_\text{Clb}$) & 0.089 & 0.039 &

$B$ & ($p_\text{Kr}$,$p_\text{Clb}$) & 0.032 & 0.031 \\

 $B$ & ($p_\text{Kr}$,$p_\text{Ks}$) & 0.029 & 0.021 &

$B$ & ($p_\text{Ks}$,$p_\text{Cap}$) & 0.024& 0.016 \\

$B$ & ($p_\text{Ks}$,$p_\text{ClCa}$) & 0.019& 0.018 &

$B$ & ($p_\text{Ks}$,$p_\text{ClCa}$) & 0.019& 0.018 \\

 $B$ & ($p_\text{Ks}$,$p_\text{Kp}$) & 0.023& 0.016 &

 $B$ & ($p_\text{Ks}$,$p_\text{Na}$) & 0.023& 0.017 \\

$B$ & ($p_\text{Ks}$,$p_\text{Nab}$) & 0.022& 0.017 &

$B$ & ($p_\text{Ks}$,$p_\text{tof}$) & 0.024& 0.017 \\

 $B$ & ($p_\text{Kp}$,$p_\text{Na}$) & 0.0087& 0.0066 &

 $B$ & ($p_\text{Kp}$,$p_\text{Nab}$) & 0.0084& 0.0063 \\

 $B$ & ($p_\text{Kp}$,$p_\text{tof}$) & 0.0098& 0.0065 &

 $B$ & ($p_\text{Na}$,$p_\text{tof}$) & 0.0055& 0.0052 \\

$V_\text{rest}$   & ($p_\text{Kr}$,$p_\text{Clb}$) & 0.11 & 0.059 &

 $V_\text{rest}$   & ($p_\text{NaCa}$,$p_\text{ClCa}$) & 0.026 & 0.024 \\

 $V_\text{rest}$   & ($p_\text{NaCa}$,$p_\text{Kp}$) & 0.024 & 0.022 &

 $V_\text{max}$   & ($p_\text{K1}$,$p_\text{Clb}$) & 0.028 & 0.019 \\
 \bottomrule
\end{tabular}
\caption{Selected second-order sensitivity indices of 18 parameter pairs for 4 AP biomarkers. $S^k_{ij}$ is the second Sobol index given by equation
\eqref{eq:index} which measures the pairwise interaction influence of
the parameters $p_i$ and $p_j$ on a single AP biomarker $y^k$. Similar with the work by \cite{nossent2011sobol}, only the parameter pairs with positive lower limit of the CI are included, showing their significant influences on AP biomarkers.\label{tab:2nd}}
\end{center}
\end{table}

\begin{figure}[t]
\centering
\includegraphics[width=1.0\textwidth]{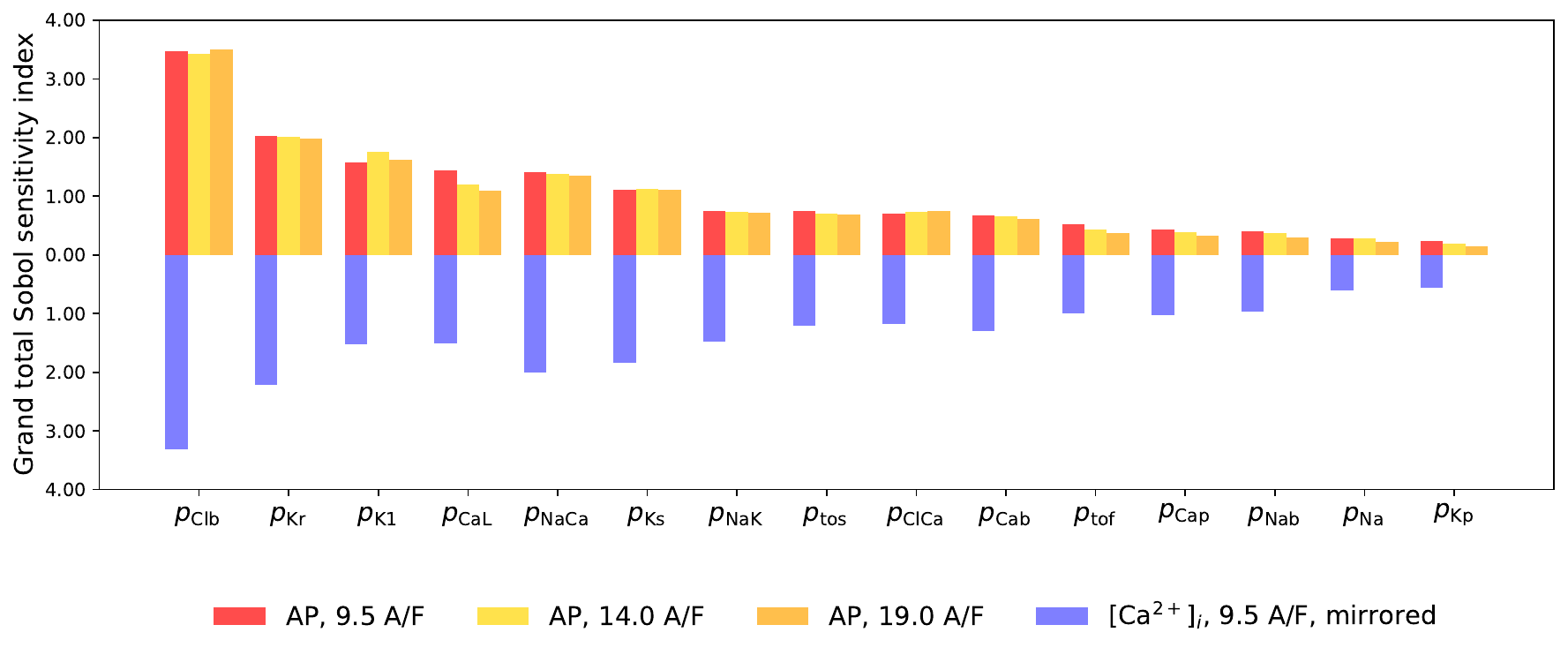}
 \caption{\black{Global sensitivity ranking of the parameters  considered in the analysis. 
Barplots of the grand total Sobol sensitivity index with respect to AP biomarkers for three different stimulus amplitudes, as well as with respect to $[\text{Ca}^{2+}]_{i}$ biomarkers (for amplitude 9.5 A/F only) are included as indicated in the legend.} \label{fig040}}
\end{figure}

\subsection{The most and the least important ionic currents in Shannon's model}
%\subsection{The grand total-order, first-order and total sensitivity index}
\label{section:discuss}

%{Grand total index
We are now ready to rank the relative strengths of ionic currents in
the Shannon model based on their influence on the model output
biomarkers. The relative strengths of ionic currents \( I_i \),
compared to their baseline values from
\citep{shannon2004mathematical}, are represented by the factors \( p_i
\), while the output biomarkers are listed in Table \ref{tab:AP
  feature}. The importance of each current is quantified using the
grand total Sobol sensitivity index, defined in equation
\eqref{eq:grand_total}, which measures the combined effect of a single
parameter \( p_i \) on all biomarkers. The grand total Sobol indices
for all 15 currents are shown in Fig \ref{fig040}, ranked in
decreasing order of their influence on model outputs. We find that, in the
notation of equation \eqref{eq:ordered} and for stimulus amplitude 9.5 A/F, the vector of ordered parameters is
\begin{equation}
    \label{ranking}
\langle p_\text{Clb}, p_\text{Kr}, p_\text{K1}, p_\text{CaL}, p_\text{NaCa}, p_\text{Ks}, p_\text{NaK}, p_\text{tos}, p_\text{ClCa}, p_\text{Cab}, p_\text{tof}, p_\text{Cap}, p_\text{Nab}, p_\text{Na}, p_\text{Kp} \rangle.
\end{equation}
\black{Increasing the stimulus amplitude up to 2 times the threshold value does not significantly affect the ranking as also shown in Fig \ref{fig040}.}
The relative strength of the background chloride current \(
p_{\text{Clb}} \), emerges as the most influential parameter in the
Shannon model, followed by that of the potassium current \(
p_{\text{Kr}} \), the inward rectifier potassium current \(
p_{\text{K1}} \), the L-type calcium current \( p_{\text{CaL}} \), the
sodium-calcium exchanger current \( p_{\text{NaCa}} \), and the slow
delayed rectifier potassium current \( p_{\text{Ks}} \). On the other
hand, the relative strength of the background potassium current \(
p_{\text{Kp}} \) is found to be the least influential parameter in the
model. This rank list constitutes the central result of our analysis. 

% Consistency with the cardiac electrophysiological research
The results summarised in Fig \ref{fig040} are in general agreement with
published experimental findings in terms of identifying the major
conductances that influence repolarization of the cardiac AP. A surprising observation emerged from the ranking of global
sensitivity is that the highest sensitivity is attributed to the value of the 
conductance of the background chloride current ($I_\text{Clb}$). \black{While this was not expected, a large sensitivity to $I_\text{Clb}$ was also independently reported in the work of \cite{Krogh2011} for the Grandi human atrial AP model \citep{Grandi2011} which in turn is based on the Shannon model.  
One possible explanation may be that because the intracellular concentration of Cl is constant in the definition of the Shannon model
the alteration of this current does not influence this concentration in simulations, but in reality this concentration could be significantly altered when varying $I_\text{Clb}$. Our sensitivity analysis reflects the behaviour of the model as currently formulated.}
\black{The importance of $I_\text{Clb}$ seems to be generally underappreciated as noted by \cite{Duan2009} despite its
importance in determining AP duration and the resting
membrane potential, both key parameters determining the electrical
stability of the heart \citep{hiraoka1998role}.} In contrast, the next most influential current, the
rapidly inactivating delayed rectifier potassium current ($I_\text{Kr}$),
carried by the hERG channel, is considerably more intensively studied
due to the linkage to the pro-arrhythmic condition associated with
long QT. The activation of $I_\text{Kr}$ current late in the repolarization
phase ensures the rapid restoration of the resting membrane potential \citep{Carmeliet1993,Varr1993} and reduced $I_\text{Kr}$ prolonges the
AP and therefore the QT duration. The slowly inactivating
delayed rectifier potassium current ($I_\text{Ks}$) contributes to the cardiac
repolarization, particularly during increased heart rates, by
preventing excessive prolongation of the AP \citep{Wu2020}. The inwardly rectifying potassium
current ($I_\text{K1}$) is important in determining the resting membrane
potential, the initial depolarization and the final repolarization
phases of the AP \citep{dhamoon2005inward,Shimoni1992}. Early repolarization (phase 1) is shaped by the
transient outward potassium current ($I_\text{tos}$), which contributes to the
characteristic notch in the AP \citep{Niwa2010}. The L-type calcium current ($I_\text{CaL}$), contributes to the initial
depolarization (phase 0) initiated by the activation of inward sodium
current by contributing to the maximum positive potential \citep{Linz2000}. Sustained inward calcium current during the plateau of
the AP (phase 2) helps to maintain depolarized
potentials and therefore the overall AP duration \citep{Linz2000}. The
magnitude and direction of the sodium-calcium exchanger current ($I_\text{NaCa}$)
is dependent on the intracellular sodium and calcium concentrations
and the transmembrane voltage \citep{Shattock2015}. At peak systolic calcium, the high
sub-sarcolemma calcium concentrations ensures that the $I_\text{NaCa}$ is an
inward current, causing net efflux of calcium from the cell and
contributing to the plateau phase of the AP.  On
repolarization due to activation of delayed rectifier currents, $I_\text{NaCa}$
remains an inward current and is one of the main calcium efflux
mechanisms \citep{Weber2002}. The sodium-potassium pump
current ($I_\text{NaK}$) is a consequence of the electrogenic stoichiometry of
the pump that actively extrudes sodium while importing potassium,
maintaining intracellular ion homeostasis. The $I_\text{NaK}$ magnitude is
normally small compared to the other currents listed above and
therefore the direct influence on the AP is small. But
the activity of the sodium potassium pump maintains normal
intracellular sodium levels and thereby indirectly influences
intracellular calcium via the sodium calcium exchanger \citep{Shattock2015}. The indirect effects on calcium sensitive currents and
the effects of altered intracellular potassium and sodium
concentrations on ionic currents explain the significant indirect
influence of the sodium potassium pump on the AP waveform \citep{Britton2017}.

\begin{figure}[t]
\centering
\includegraphics[width=0.92\textwidth, height=0.845\textheight]{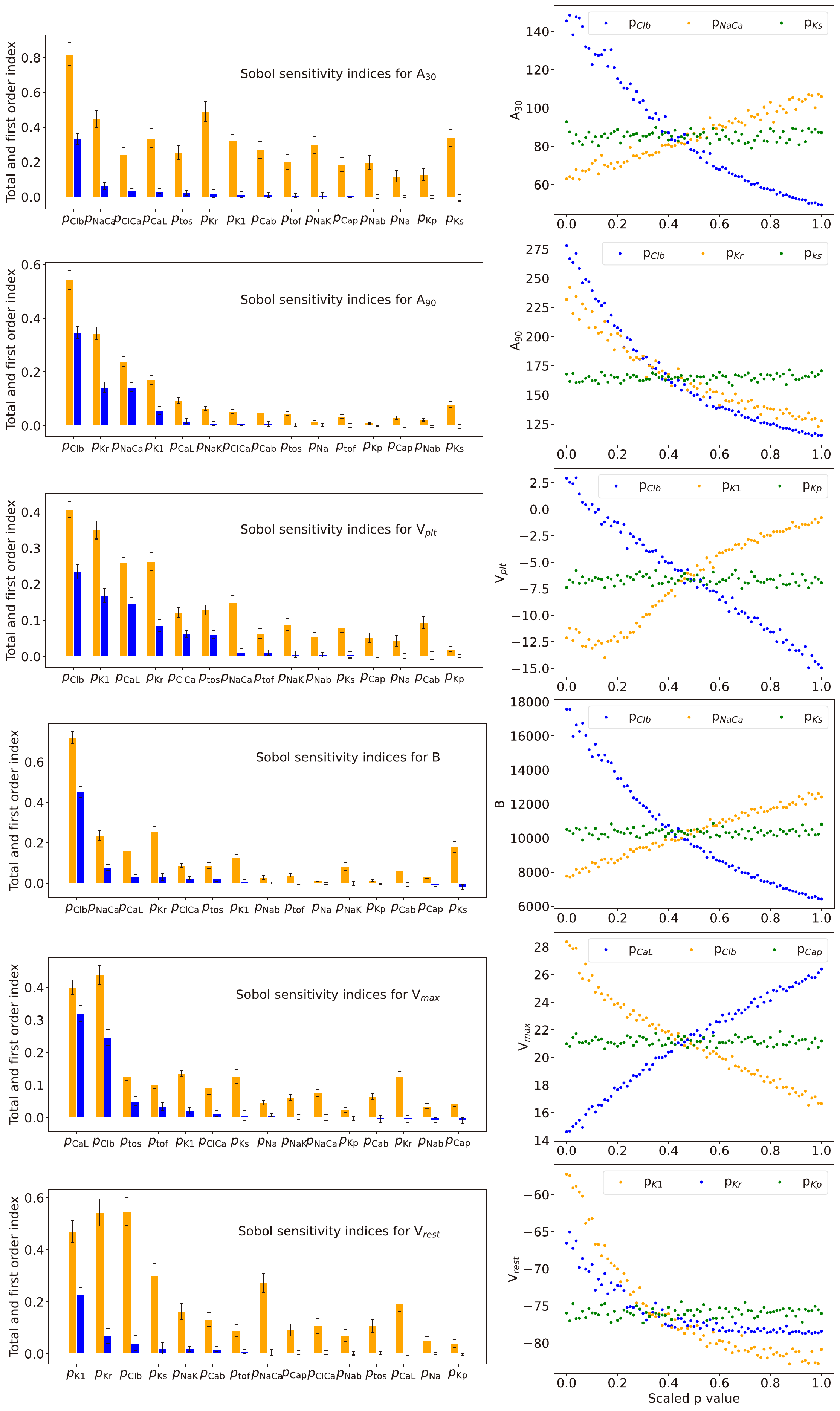}
 \caption{
(left column) Total and first-order Sobol sensitivity
    indices for the six studied biomarkers sorted in descending order
    with respect to the first-order index. (right column) Conditional
    expectations for the parameters with the largest, the
    second-largest and the smallest values of the first-order Sobol
    sensitivity index. To aid visualisation parameter values $p$
      are scaled by ${p}_\text{scaled} =
      ({p-{p}_\text{min}})/({{p}_\text{max}-{p}_\text{min}})$ to
      $[0,1]$ on the abscissa.}  \label{fig050}
\end{figure}

% Total index: 
To produce a secondary ranking of the ionic currents based on their
influence on individual biomarkers, we analyze the corresponding
total, first-order, and second-order indices. The total and
first-order indices are shown in the left columns of Fig
\ref{fig050}. 
% Say in plain text what the total index is
The total-order index, defined in Equation \eqref{eq:total}, quantifies a
parameter's contribution to the variability of a single biomarker,
accounting for both direct impacts and interactions with other
parameters. 
% describe and interpret results 
% refer to p's with their full names not only with their symbols.
The background chloride current relative strength parameter, \(
p_{\text{Clb}} \), is consistently the most influential for all action
potential biomarkers. Following \( p_{\text{Clb}} \), the rapid
delayed rectifier potassium current parameter (\( p_{\text{Kr}} \)),
the inward rectifier potassium current parameter (\( p_{\text{K1}}
\)), the sodium-calcium exchange current parameter (\( p_{\text{NaCa}}
\)), and the L-type calcium current parameter (\( p_{\text{CaL}} \))
exhibit significant total-order sensitivity. 

% First-order index 
To determine whether specific ionic currents influence a biomarker
individually or through interactions with other currents, we analyze
their first-order Sobol indices. Defined in Equation
\eqref{eq:indices}, the first-order index quantifies the direct effect
of a single parameter on a single biomarker, representing the
proportion of variance caused solely by that parameter. 
The first-order indices for the relative strengths of ionic currents
are shown in the left column of Fig \ref{fig050}.
%The channel responsible for the background current are thought be to distinct from the Ca$^{2+}$ activated chloride conductance \citep{Duan2009}. 
The background chloride current parameter (\( p_{\text{Clb}} \)) is
consistently dominant, particularly for biomarkers such as action
potential duration at 30\% and 90\% repolarization (\( A_{30} \) and
\( A_{90} \)), plateau voltage, and AP bulk. For
example, \( p_{\text{Clb}} \) contributes five times more to the
variance of \( A_{30} \) than the second- and third-ranked
parameters. 
The resting membrane potential (\( V_\text{rest} \)) is most sensitive
to the potassium current parameters (\( p_{\text{K1}} \) and \(
p_{\text{Kr}} \)), while the maximum AP upstroke (\(
V_\text{max} \)) is strongly influenced by the L-type calcium current
parameter (\( p_{\text{CaL}} \)) as it is the major depolarizing current during late upstroke. This effect of calcium current is through the effect on (\(
V_\text{max} \)), not influencing the maximum upstroke velocity as (\( p_{\text{Na}} \)) is almost the sole determinant parameter \citep{Shaw1997}. The sodium-calcium exchange current
parameter (\( p_{\text{NaCa}} \)) consistently ranks among the top
three for several biomarkers, highlighting its importance in action
potential dynamics. Conversely, the background potassium current
parameter (\( p_{\text{Kp}} \)) and the slow delayed rectifier
potassium current parameter (\( p_{\text{Ks}} \)) exhibit the lowest
first-order indices across all biomarkers.

% second-order sensitivity index
To assess the joint influence of two currents on biomarkers, we
examine their second-order Sobol indices. Table \ref{tab:2nd} lists
all second-order indices with a positive lower confidence interval
limit. Notably, \( S_{\text{Kr},\text{Clb}} \) emerges as the highest
second-order index for \( A_{90} \), \( B \), and \( V_\text{rest} \),
while \( S_{\text{K1},\text{Clb}} \) dominates for \( V_\text{max} \).  
The strong interaction between Cl\(^{-}\) and K\(^{+}\) conductances
aligns with findings in \citep{hiraoka1998role}, which suggest that
the role of \( I_\text{Clb} \) in AP waveform variability depends on
its interplay with other ion conductances, especially
K\(^{+}\). Reducing extracellular potassium concentration to suppress
\( I_\text{Kr} \) and \( I_\text{K1} \) increases resting potential
depolarization, amplifying the depolarizing action of \( I_\text{Clb}
\) \citep{hiraoka1998role}. 
Additionally, the slow delayed rectifier potassium current parameter
(\( p_{\text{Ks}} \)) shows significant second-order interaction
effects on the AP bulk (\( B \)), particularly with parameter \(
p_{\text{Kr}} \). This supports findings by
\cite{jost2005restricting} that blocking \( I_\text{Ks} \) markedly
prolongs AP duration when “repolarization reserve” is reduced by \(
I_\text{Kr} \) block.

% Relationships between AP biomarkers and parameters
Having identified the parameters that most strongly influence the AP
biomarkers, we next explore the explicit relationships between
them. The second column of Fig \ref{fig050} displays the
conditional expectations of the biomarkers with respect to the
parameters with the highest and lowest first-order Sobol
indices. These plots show the expected value of a biomarker as a
function of a parameter while all other parameters vary
randomly. These conditional expectations provide insights into the
general dependence of biomarkers on specific parameters and indicate
how the biomarker is likely to vary as one parameter changes while the
values of all others remain uncertain. 
%The corresponding marginal
%distributions are provided in \black{Supplementary Figure YYY ref{fig:GRfull}}. 

% Comparison with prior research works
The work of \citet{romero2011systematic} highlights the
significant influence of repolarization currents, particularly \(
p_\text{CaL} \), \( p_\text{Kr} \), \( p_\text{tos} \), \(
p_\text{NaK} \), and \( p_\text{NaCa} \), on AP waveform. Consistent
with this, we find these parameters ranked within the top eight in
Fig \ref{fig040}.  
Unlike \citet{romero2011systematic}, our analysis reveals \(
p_\text{Ks} \) as influential, likely due to differences in
methodology. Their local sensitivity method varied selected parameters
by only \(\pm15\%\) and \(\pm30\%\) from the Shannon model's baseline
values \citep{shannon2004mathematical}, limiting the analysis to a
narrow parameter neighborhood and ignoring interaction effects. In
contrast, our global sensitivity method explores a wide parameter
space, allowing all parameters to vary simultaneously. This broader
approach captures interaction effects, as illustrated in Fig
\ref{fig050}, where \( G_\text{Ks} \) exhibits low first-order
indices for six AP biomarkers but high total-order indices, indicating
strong higher-order interaction effects on the AP waveform. 

The work of \citet{gemmell2014population} systematically explored the
effects of simultaneously varying the magnitude of six transmembrane
current conductances in the \citet{shannon2004mathematical}
model. Using clutter-based dimension reordering, they identified \(
p_\text{CaL} \) as having the greatest influence on AP variability at
both 400 ms and 1000 ms, along with \( p_\text{K1} \) and \(
p_\text{to} \) at 400 ms and 1000 ms, respectively. These findings
align with our results, where these three parameters also ranked among
the top eight in Fig \ref{fig040}, despite our analysis being
conducted at a different basic cycle length. 

In contrast to the findings of \cite{coveney2020sensitivity}, \(
p_{\text{Na}} \) is not the most influential parameter in this
study. This discrepancy may stem from our focus on successfully
excited cells, which \black{for stimulus amplitude 9.5 A/F} limits the range of \( p_{\text{Na}} \) to values
\( >1 \). \black{If  values of $p_\text{Na} < 1$ were to be considered, the effects would have included the differences between normal and failing AP responses which are, of course, huge. The effects of $p_\text{Na}$ on normal APs alone are far less significant.}

\subsection{Extension sensitivity analysis for intracellular calcium biomarkers}
\black{A number of studies using populations of models use biomarkers of intracellular calcium concentration to calibrate them \citep{Varshneya2021,Llopis-Lorente2023}. To facilitate this, we have extended the Sobol sensitivity analysis to include the following four intracellular calcium biomarkers:
(a) ED:  $[\text{Ca}^{2+}]_{i}$ at the end of diastole; (b) PSV: peak systolic value of $[\text{Ca}^{2+}]_{i}$; (c) $\text{T}_\text{peak}$: time from stimulus to peak $[\text{Ca}^{2+}]_{i}$; (d) $\text{D}_{50}$: period of time when $[\text{Ca}^{2+}]_{i}$ remains elevated above a threshold of 50\% recovery from the peak value to the resting value, informally duration at 50\% amplitude. The grand total Sobol indices based on these four biomarkers are shown in blue in Fig \ref{fig040} for the standard setup of our study. The top seven most influential parameters remain the same as those found by analysis of AP-biomarkers shown in red in Fig \ref{fig040}, with the relative strength of the background chloride current $p_{\text{Clb}}$ still emerging as the most influential parameter, followed by $p_{\text{Kr}}, p_{\text{NaCa}}, p_{\text{Ks}}$. On the other hand, the relative strength of the background potassium current $p_{\text{Kp}}$ remains to be the least influential parameter in the model. The higher rank of $p_{\text{NaCa}}$ reflects its large importance for intracellular calcium dynamics and this is because the rise in intracellular calcium concentration activates the Na+/Ca2+ exchanger cell membrane pump \citep{Bogdanov2001}.}

%\begin{figure}[t]
%\centering
% \includegraphics[width=0.9\textwidth]{grand_total_Cai.pdf}
%  \caption{Global sensitivity ranking of the parameters considered in the analysis for intracellular calcium biomarkers. Barplots of the grand
%total Sobol sensitivity index of the fifteen maximal current density parameters sorted in decreasing
%value.. \label{fig_Cai}} 
%\end{figure}

\subsection{A hierarchy of reduced Shannon models}
%\subsection{The test of reducing the parameter space dimension}
\label{section:SA repeat}
Using the ranked Shannon model parameters based on their grand Sobol
sensitivity indices (see Fig \ref{fig040} and equation
\eqref{ranking}), we now address the second main goal of our
sensitivity analysis: constructing a hierarchy of reduced Shannon
models. This hierarchy is formed by successively fixing the least
significant parameters to their baseline values, as follows: 
\begin{align}
    \label{hierarchy}
    \mathbf{y} &= \mathbf{f}\big(\langle p_\text{Clb}, p_\text{Kr}, p_\text{K1}, p_\text{CaL}, p_\text{NaCa}, p_\text{Ks}, p_\text{NaK}, p_\text{tos}, p_\text{ClCa}, p_\text{Cab}, p_\text{tof}, p_\text{Cap}, p_\text{Nab}, p_\text{Na}, p_\text{Kp}\rangle\big), \nonumber\\
    \mathbf{y}^{\langle 14\rangle} &= \mathbf{f}\big(\langle p_\text{Clb}, p_\text{Kr}, p_\text{K1}, p_\text{CaL}, p_\text{NaCa}, p_\text{Ks}, p_\text{NaK}, p_\text{tos}, p_\text{ClCa}, p_\text{Cab}, p_\text{tof}, p_\text{Cap}, p_\text{Nab}, p_\text{Na}, 1 \rangle\big),\nonumber\\
    \mathbf{y}^{\langle 13\rangle} &= \mathbf{f}\big(\langle p_\text{Clb}, p_\text{Kr}, p_\text{K1}, p_\text{CaL}, p_\text{NaCa}, p_\text{Ks}, p_\text{NaK}, p_\text{tos}, p_\text{ClCa}, p_\text{Cab}, p_\text{tof}, p_\text{Cap}, p_\text{Nab}, 1, 1 \rangle\big),\nonumber\\
    & \cdots  \\
    \mathbf{y}^{\langle 1\rangle} &= \mathbf{f}\big(\langle p_\text{Clb}, 1,1,1,1,1,1,1,1,1,1,1,1,1,1\rangle\big).\nonumber
\end{align}

\black{The original Shannon model \citep{shannon2004mathematical} corresponds to
\begin{align}
\mathbf{y}^{\langle 0\rangle} &= \mathbf{f}\big(\langle 1, 1,1,1,1,1,1,1,1,1,1,1,1,1,1\rangle\big),
\end{align}
representing the lowest hierarchical level where all parameters are fixed.}

\begin{figure}[t]
\centering
\includegraphics[width=0.9\textwidth]{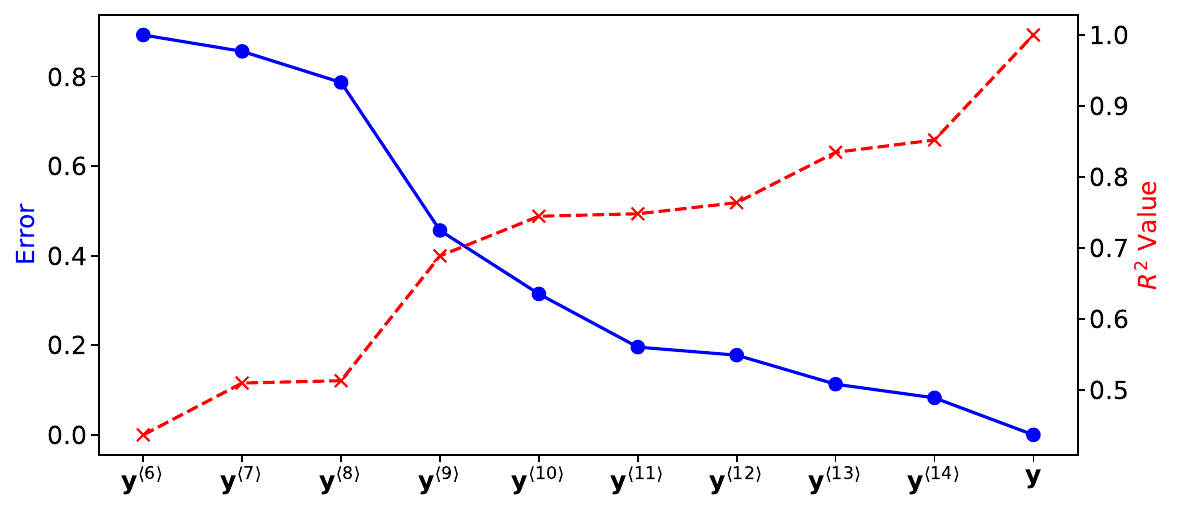}
  \caption{Deviation of reduced models $\mathbf{y}^{\langle M
      \rangle}$ from the full model $\mathbf{y}$ as measured by the
    mean relative error (blue dots; left ordinate axis) and the
    coefficient of determination $R^2$ (red crosses, right ordinate
    axis). \label{fig060}} 
\end{figure}

\begin{figure}[t]
\centering
\includegraphics[width=1.0\textwidth]{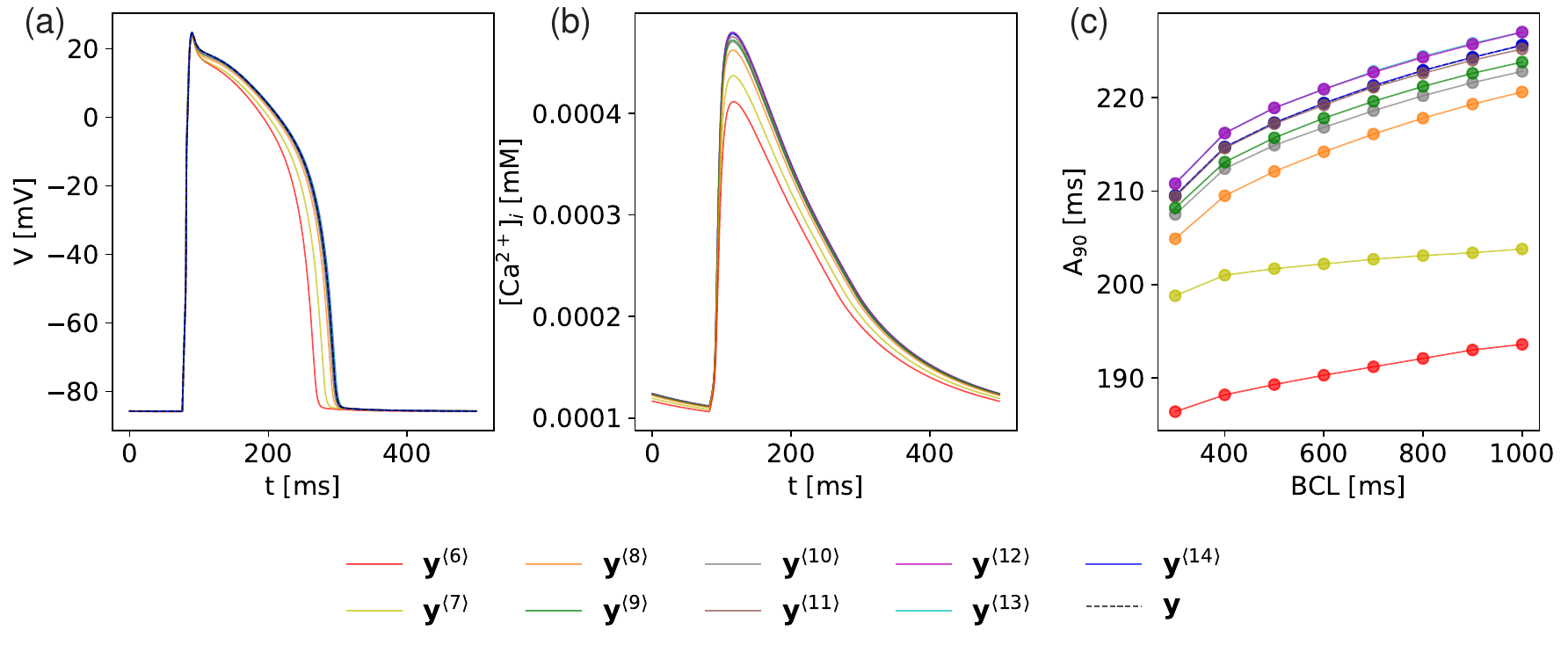}
  \caption{A local illustration of the accuracy of reduced models. Panel (a) and (b) show AP forms, and Calcuim transients $[\text{Ca}]^{2+}_{i}$, while panel (c) shows $A_{90}$ restitution curves ($A_{90}$ for periodically excited waveforms as a function of basic cycle length) for reduced models $\mathbf{y}^{\langle 6 \rangle}$ to $\mathbf{y}^{\langle 14 \rangle}$ in comparison with the full model $\mathbf{y}$. The models are evaluated locally so that for the full model all parameters are set equal to 0.8, while for the models at hierarchical level $M$ the $M$ "variable" parameters are also set to 0.8 while the rest are kept at baseline value corresponding to $1$.   
  \label{fig070}}
\end{figure}

Fig \ref{fig060} quantifies the global discrepancy
between the full model (\( \mathbf{y} \)) and the reduced models (\(
\mathbf{y}^{\langle M \rangle} \)) at hierarchical levels \( M \) over
the parameter ranges of normal response. Two measures of discrepancy
are presented: the mean relative error, defined by Equation
\eqref{eq:error} and plotted on the left-hand ordinate axis, and the
coefficient of determination (\( R^2 \)), shown on the right-hand
ordinate axis, both as functions of \( M \).
For models \( \mathbf{y}^{\langle 0 \rangle} \) to \(
\mathbf{y}^{\langle 5 \rangle} \), where fewer than the six most
sensitive parameters are varied, the discrepancy is non-monotonic and
remains too large. Consequently, these models do not approximate the
full model well and are excluded from the plots in Fig
\ref{fig060}, which only includes models from \(
\mathbf{y}^{\langle 6 \rangle} \) to \( \mathbf{y} \).  
At higher hierarchical levels (\( M \geq 6 \)), the discrepancy
decreases monotonically with increasing \( M \). The mean relative
error approaches zero, while \( R^2 \) approaches unity, indicating
improved accuracy. Fig \ref{fig070} provides a
local comparison of the reduced and full models to offer an intuitive
understanding of their accuracy. 

\begin{figure}[t]
\centering
\includegraphics[width=\textwidth]{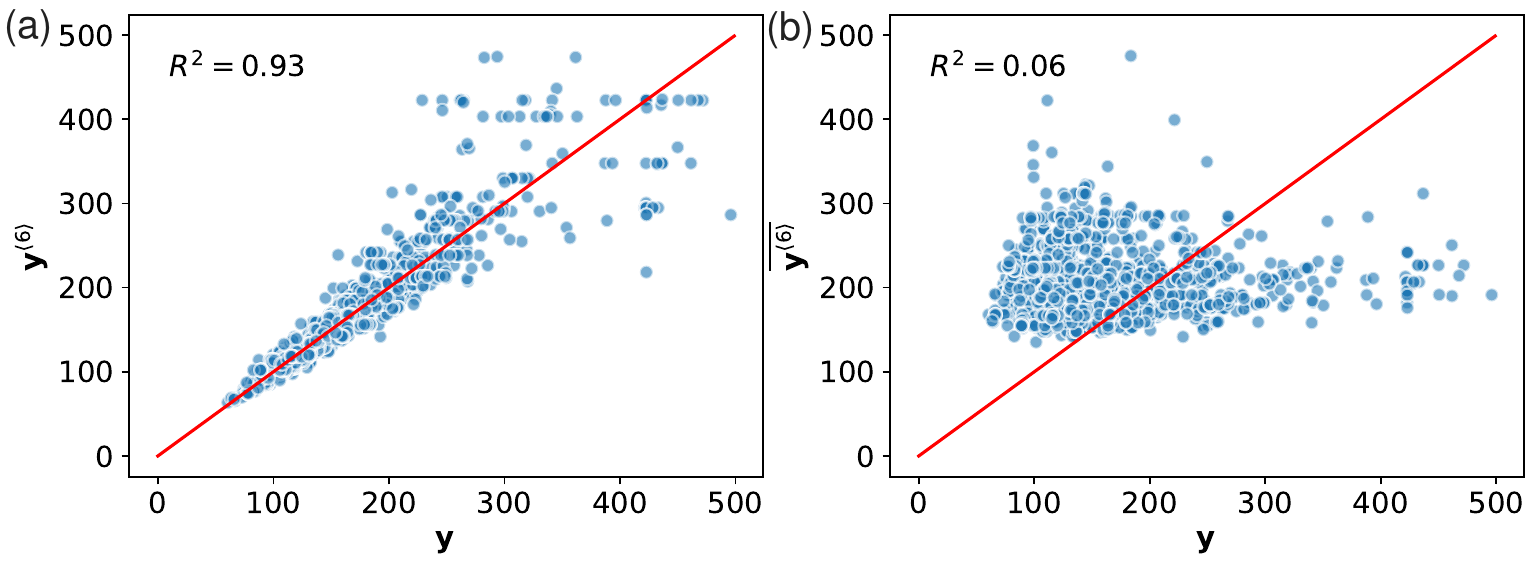}
 \caption{Scatter plots of $A_{90}$ values obtained from the reduced model $\mathbf{y}^{\langle  6\rangle}$ (a) and the reduced model $\overline{\mathbf{y}^{\langle  6\rangle}}$ (b) compared to the full model $\mathbf{y}$. 
 \label{fig080}}
\end{figure}

While a hierarchy of models can be constructed by ranking parameters
based on their grand total sensitivity index, more “economical” specialized reduced models can be obtained to capture individual
biomarkers. For each biomarker of interest, this is achieved by
ranking parameters based on their total sensitivity indices.  
To illustrate, Fig \ref{fig080} presents scatter plots of \(
A_{90} \) values obtained from the reduced model 
\[
\mathbf{y}^{\langle 6 \rangle} = \mathbf{f}\big(\langle p_{\text{Clb}}, p_{\text{Kr}}, p_{\text{NaCa}}, p_{\text{K1}}, p_{\text{CaL}}, p_{\text{Ks}}, 1,1,1,1,1,1,1,1,1 \rangle\big),
\]
and its “complement” model
\[
\overline{\mathbf{y}^{\langle 6 \rangle}} = \mathbf{f}\big(\langle 1,1,1,1,1,1, p_{\text{NaK}}, p_{\text{tos}}, p_{\text{ClCa}}, p_{\text{Cab}}, p_{\text{tof}}, p_{\text{Cap}}, p_{\text{Nab}}, p_{\text{Na}}, p_{\text{Kp}} \rangle\big),
\]
against the corresponding values from the full model in panels (a) and
(b), respectively. 
The coefficient of determination (\( R^2 \)) between the reduced model
(\( \mathbf{y}^{\langle 6 \rangle} \)) and the full model (\(
\mathbf{y} \)) is 0.93, while \( R^2 \) for the complement model (\(
\overline{\mathbf{y}^{\langle 6 \rangle}} \)) and the full model is
only 0.06. This demonstrates that the reduced model \(
\mathbf{y}^{\langle 6 \rangle} \), while insufficient to capture all
six biomarkers simultaneously, accurately reproduces the single
biomarker \( A_{90} \).  
Table \ref{tab:R value} lists other specialized reduced models that
capture each of the six biomarkers with \( R^2 \geq 0.9 \), a
threshold often considered sufficient to assess how well one variable
replicates another. 

\begin{table}[t]%%%Table caption goes here
\begin{center}
\begin{tabular}{lll}%%%The number of columns has to be defined here
\hline
AP biomakers & $R^2$ value &Parameters \\
\hline
$A_{90}$ &0.93 &$p_{\text{Clb}}, p_{\text{Kr}}, p_{\text{NaCa}}, p_{\text{K1}}, p_{\text{CaL}}, p_{\text{Ks}}$\\
$A_{30}$     &  1.0        &  $p_{\text{Clb}}, p_{\text{Kr}}, p_{\text{NaCa}}, p_{\text{Ks}}, p_{\text{CaL}}, p_{\text{K1}}, p_{\text{NaK}}, p_{\text{Cab}}, p_{\text{tos}}, p_{\text{ClCa}}$  \\
 &  &  $p_{\text{tof}}, p_{\text{Nab}}, p_{\text{Cap}}, p_{\text{Kp}}, p_{\text{Na}}$\\
$V_\text{max}$ &0.91 &$p_{\text{Clb}}, p_{\text{CaL}}, p_{\text{K1}}, p_{\text{Ks}}, p_{\text{Kr}}, p_{\text{tos}}, p_{\text{tof}}, p_{\text{ClCa}}, p_{\text{NaCa}}, p_{\text{Cab}},p_{\text{NaK}}$ \\
$V_\text{rest}$ &0.92 & $p_{\text{Clb}}, p_{\text{Kr}}, p_{\text{K1}}, p_{\text{Ks}}, p_{\text{NaCa}}, p_{\text{CaL}}, p_{\text{NaK}}, p_{\text{Cab}},p_{\text{tos}},p_{\text{ClCa}},$ \\
      &           & $p_{\text{Cap}}, p_{\text{tof}}, p_{\text{Nab}}, p_{\text{Na}}$  \\
$B$ &0.91 &$p_{\text{Clb}}, p_{\text{Kr}}, p_{\text{NaCa}}, p_{\text{Ks}}, p_{\text{CaL}}, p_{\text{K1}}, p_{\text{ClCa}}, p_{\text{tos}}$\\
$V_\text{plt}$ &0.92 & $p_{\text{Clb}}, p_{\text{K1}}, p_{\text{Kr}}, p_{\text{CaL}}, p_{\text{NaCa}}, p_{\text{tos}}, p_{\text{ClCa}}, p_{\text{Cab}}, p_{\text{NaK}}, p_{\text{Ks}}$ \\\hline
\end{tabular}
\caption{Minimal sets of parameters for specialised reduced models that capture a specific biomarker with
a coefficient of determination value $R^2$ of at least 0.9. Parameters are ranked by their total-order index from large to small. \label{tab:R value} All 15 parameters are required for $A_{30}$ to achieve high accuracy (\( R^2 \geq 0.9
\)) because of their non-negligible high-order interaction effects.}
\end{center}
\end{table}

The appropriate accuracy of a Shannon model reduction depends on
the specific application is intended to investigate. The
discrepancy analysis presented in this section provides a framework
for selecting a hierarchical model that balances simplicity and
accuracy to meet the requirements of the application.

\section{Conclusion}
\label{section:conclude}
% Summary 
This study presents a global sensitivity analysis of the
Shannon model of rabbit ventricular myocyte electrophysiology,
focusing on the influence of ionic current strength parameters on key
AP biomarkers. By ranking parameters using Sobol sensitivity
indices, we have identified the most influential ionic currents, including
\( p_{\text{Clb}}, p_{\text{Kr}}, p_{\text{K1}}, p_{\text{CaL}}, p_{\text{NaCa}}\)
and \( p_{\text{Ks}} \), which dominate the model's variability.  

Our analysis demonstrates the utility of a global sensitivity approach
in capturing both direct effects and interaction effects among
parameters, providing a deeper understanding of their role in shaping
AP dynamics. For instance, \( p_{\text{Ks}} \), which
exhibited low first-order indices but high total-order indices, was
found to contribute significantly through interaction effects,
highlighting the importance of considering higher-order sensitivity
metrics. 

The hierarchical model framework constructed in this study offers a
practical approach to reducing model complexity while maintaining
accuracy. Models that exclude less influential parameters (\( M \geq 8
\)) were shown to adequately approximate the full model, as evidenced by
monotonic reductions in mean relative error and increases in \( R^2
\). Additionally, specialized reduced models targeting individual
biomarkers were developed, achieving high accuracy (\( R^2 \geq 0.9
\)) while requiring fewer parameters.
The findings provide a foundation for tailoring the Shannon model to
other specific applications, balancing simplicity with accuracy. The
insights gained into parameter importance and interaction effects can 
guide future investigations, including the development of reduced
models for multi-scale simulations and the exploration of
electrophysiological variability across different cell types and
conditions.  
As an example, in our prior work on inter-cell variability of rabbit
ventricular electrophysiology \citep{Lachaud2022,
  simitev2025largepopulationcellspecificaction} parameter choices were made by educated
guesses. The 
results of the current analysis complement these works and provide
justification for their parameter choices.

%{Limitations, Extensions, and Future Directions}
A notable limitation of this study is its focus on sensitivity
analysis at a fixed stimulation rate of 2 Hz, without systematically
exploring AP dynamics at other pacing rates. This narrow
scope may limit the generalizability of the findings across diverse
physiological and pathological conditions where cellular pacing rates
vary. Future studies should extend this analysis to include a broader
range of stimulation frequencies to enhance
applicability. Additionally, the choice and number of model parameters
and biomarkers warrant further investigation. While this study focused
on a subset of parameters, the Shannon model includes nearly 200
parameters, and incorporating biomarkers such as restitution
properties or AP duration rate adaptation could provide a more
comprehensive perspective. However, currently there are no experimental studies that assess the intracellular calcium, restitution, or other biomarkers in addition to AP biomarkers. 
Expanding the sensitivity analysis to
alternative AP models, such as the \citet{Mahajan2008}
model for rabbit ventricular myocytes, would also enhance the
robustness of the findings. \black{Our sensitivity analysis reflects the behavior of the Shannon model as currently formulated. However, the assumption of constant intracellular chloride concentration in the model is a simplification that may exaggerate the influence of \( p_{\text{Clb}}\). Future extensions incorporating dynamic chloride ion homeostasis will be crucial to reassessing the physiological relevance of this finding. Nonetheless, the result highlights a potentially unappreciated role of chloride currents in AP dynamics, meriting further experimental and modelling investigation.}

From a technical standpoint, alternative sensitivity analysis methods
offer potential avenues for improving 
accuracy and efficiency. Comparative studies could evaluate whether
these methods yield better insights or computational benefits. The
parameter space dimension reduction proposed in this study provides
valuable guidance for selecting fitting parameters in sample-specific
electrophysiological modeling, enabling the study of specific cellular
states and predictions of cellular behavior under varied external
conditions. 

%A promising extension involves implementing the method developed by
%\citet{lazarus2022sensitivity} to apply inverse uncertainty
%quantification to the most influential model parameters. This
%approach could evaluate parameter identifiability and uniqueness,
%contributing to a deeper understanding of model reliability and
%predictive power. 

\begin{plain}
\section*{Statements}
\newcommand{\ethics}[1]{\paragraph*{Ethics.} #1}
\newcommand{\dataccess}[1]{\paragraph*{Data availability statement.} #1}
\newcommand{\competing}[1]{\paragraph*{Conflict of interest.} #1}
\newcommand{\funding}[1]{\paragraph*{Funding.} #1}
\end{plain}

%\ack{Insert acknowledgment text here.}
\funding{This work was supported by the UK Engineering and Physical Sciences
  Research Council [grant numbers EP/S030875/1 and EP/T017899/1].}

\dataccess{Data and code are
available from \url{https://github.com/ZhechaoYanggla/Sobol_paper}}.

\competing{The authors declare that no competing interests exist.}

\bibliographystyle{plos2015}

\end{document}